\documentclass[a4paper,12pt]{article}
\usepackage{amsfonts,amsthm,amsmath,amssymb,graphicx,setspace}
\usepackage{cite}
\usepackage{braket}
\usepackage{hyperref}
\usepackage{amsmath}
\usepackage{url}
\usepackage{graphicx,color}

\def\tr{{\mathrm{tr}}}

\def\sinh{{\mathrm{sinh}}}
\def\cosh{{\mathrm{cosh}}}

 \def\frac#1#2{{#1\over #2}}

\def\be{\begin{equation}}
\def\ee{\end{equation}}
\def\ba{\begin{eqnarray}}
\def\ea{\end{eqnarray}}
\numberwithin{equation}{section}

\usepackage{hyperref}
\begin{document}

\begin{titlepage}

\thispagestyle{empty}

\begin{flushright}
YITP-18-89
\end{flushright}

\bigskip

\begin{center}
\noindent{{\Large \textbf {Circuit complexity in interacting QFTs and RG flows}}}\\
\vspace{0.5cm}
	
Arpan Bhattacharyya$^{(a)}$\footnote{bhattacharyya.arpan@yahoo.com}, Arvind Shekar$^{(b)}$\footnote{arvinduniqc@gmail.com}  and Aninda Sinha$^{(b)}$\footnote{asinha@iisc.ac.in} \\ ~~~~\\
        $ {}^{(a)}$ {\it Center for Gravitational Physics, \\
\it Yukawa Institute for Theoretical Physics (YITP), Kyoto University, \\
\it Kitashirakawa Oiwakecho, Sakyo-ku, Kyoto 606-8502, Japan\\
	${}^{(b)}$\it Centre for High Energy Physics,
	\it Indian Institute of Science,\\ \it C.V. Raman Avenue, Bangalore 560012, India. \\}	
\vskip 3cm
\end{center}
\begin{abstract}

We consider circuit complexity in certain interacting scalar quantum field theories, mainly focusing on the $\phi^4$ theory. We work out the circuit complexity for evolving from a nearly Gaussian unentangled reference state to the entangled ground state of the theory. Our approach uses Nielsen's geometric method, which translates into working out the geodesic equation arising from a certain cost functional. We present a general method, making use of integral transforms, to do the required lattice sums analytically and give explicit expressions for the $d=2,3$ cases. Our method enables a study of circuit complexity in the epsilon expansion for the Wilson-Fisher fixed point. We find that with increasing dimensionality the circuit depth increases in the presence of the $\phi^4$ interaction eventually causing the perturbative calculation to breakdown. We discuss how circuit complexity relates with the renormalization group.

\end{abstract}

\end{titlepage}
\newpage
\tableofcontents

\onehalfspacing


\section{Introduction}

In the context of quantum information theory, circuit complexity is the number of unitary operations needed to perform a desired task in a quantum circuit \cite{c1,c2,c3,c4, CCintroduction,CCintroduction2, Aaronson:2016vto,watrous,osborne2012hamiltonian, gharibian2015quantum}. Of course it is desirable that the number of steps is minimum to have the most efficient implementation of a quantum algorithm. In particular it is important to quantify this so that one can have meaningful comparisons with classical algorithms. The question of circuity complexity in the context of quantum field theories is still relatively novel with very few results. In \cite{Jordan1,Jordan2, Jordan3}, it was shown that the non-perturbative calculation of $n$-particle scattering amplitude in a scalar $\phi^4$ theory on a quantum computer would have an exponential advantage over known algorithms which can be implemented on a classical computer which uses perturbative Feynman diagram techniques to perform such a calculation. This is quite remarkable and the question naturally arises how a quantum computer would compute other interesting quantities that are calculated by conventional means. Especially, is there a connection between renormalization group flows and circuit complexity?

Recently, observations  due to Susskind and collaborators based on observations related to thermalization during black hole formation in holography have spurred activity in computing  circuit complexity in quantum field theories \cite{jm,Hashimoto:2017fga,Yang:2017nfn,Khan:2018rzm,Hackl:2018ptj,Yang:2018nda,Alves:2018qfv, Magan:2018nmu,Caputa:2018kdj,Camargo:2018eof,recentmyers,ab,ab1}. In holography, it was observed that while the entanglement entropy asymptotes to a constant with time as the black hole thermalizes, the size of the Einstein-Rosen bridge in the context of the eternal AdS black hole keeps increasing. It was proposed that the analogous quantity that keeps increasing after thermalization is complexity. Two interesting proposals were given in the context of AdS/CFT \cite{Susskind1,Susskind2,Susskind3,Susskind4,Susskind5,Susskind6}. The first one is the volume of a maximal codimension-one bulk surface extending to the boundary of AdS space time (complexity = volume). This particular slice is chosen in a way such that it asymptotes to a specific time slice on which the boundary state resides. The second one is to consider the so called Wheeler-DeWitt (WDW) patch which is basically the domain of dependence of a bulk Cauchy surface anchored at a specific time. Both these two objects have the potential to probe physics behind the horizon and both of them evolve with time even after the thermal equilibrium has been reached. These two proposals have been subjected to a host of interesting checks \cite{Barbon, Alishahiha:2015rta, Hol1,Hol2,Hol3,Hol4,Hol5,Hol6,Hol7,Hol8,Hol9,Hol10,Alish,Hol11,Hol12,Hol13,Hol15,Hol16,Hol17, Moosa, Hol18,Hol19,Hol20,Hol21,Hol22,Hol23,Hol24,Hol25,Hol26, Hol14,Hol19a,Hol19b,Hol19c}.
\par
In \cite{Nielsen1,Nielsen2,Nielsen3} a geometric approach for circuit complexity was put forward which was studied in great detail in the context of free scalar field theories in \cite{jm}. It was proposed that to find the minimum complexity, one writes down a suitable cost function in the space of unitaries, which works out to be in general a Finsler space, and then minimizes it. The distance from the reference state ($|\psi^{R}\rangle$) to the target state ($|\psi^{T}\rangle$) is then the geodesic distance.  To elaborate, we want 
\begin{align}
\begin{split}
|\psi^{T}(s=1)\rangle=U(s)|\psi^{R}(s=0)\rangle,
\end{split}
\end{align}
where $U(s)$ is a unitary operator. Here $s$ parametrizes  a path in the Hilbert space with the boundary conditions such that the reference state is located at $s=0$ and the target state is located at $s=1.$ This is just a matter of convenience . We can redefine this parameter. Now every unitary operator can be written as follows,
\be
U(s)= {\overleftarrow{\mathcal{P}}}\exp(i\int_{0}^{s}\, ds\, H(s)),
\ee
where $H(s)$ is a Hermitian operator. Then $H(s)$ can be expanded in a suitable basis ($M_{I}$) in the following way, $H(s)= Y^{I}(s) M_{I}.$ $Y^{I}(s)$'s  are generally referred to  as control functions. Now given a set of these elementary gates ($M_I$) we construct our unitary $U(s).$  The basic notion of complexity (or more suitably in this case ``circuit depth") is to provide a measure to count the number of \textit{``elementary"} gates which can be combined to form the required $U(s).$ In general, this is not a unique procedure and is difficult to accomplish. Following \cite{jm,Nielsen1,Nielsen2,Nielsen3} we try to find the shortest path or the geodesic  connecting the reference and target state. To do so, we first define a suitable \textit{``cost function"} $F(U, \dot U)$  and the complexity is then defined as,
\be \label{cost}
\mathcal{D}(U)=\int_{0}^{1} \mathcal{F}(U,\dot U)\, ds.
\ee
We then minimize this cost function which then gives the  geodesic connecting the two states. Then evaluating $\mathcal{D}(U)$ on this geodesic, we obtain a measure for the complexity. Now there are various possible choices for the  \textit{``cost function"}. But there are some desirable properties \cite{Nielsen1,Nielsen2, Nielsen3} that these cost functions should satisfy.   They should be continuous, positive definite, homogeneous and satisfy the triangle inequality. These properties help us to identify these functions as legitimate functions measuring the distance between two points on the underlying manifold. In addition to these, if these functions are  infinitely differentiable, then (\ref{cost}) gives the distance between two points on a Finsler manifold.  Keeping these properties in mind we can choose various possible functions. We quote the ones advocated in \cite{jm, Hackl:2018ptj,recentmyers}
\begin{align}
\begin{split}
& \mathcal{F}_2(U,Y)=\sqrt{\sum_{I}p_{I} (Y^{I})^2}, \mathcal{F}_{\kappa}(U,Y)=\sum_{I}p_{I}\, |Y^{I}|^{\kappa}, \quad \kappa\textrm{\quad is\, an\, integer\,and } , \kappa >1,\\& \mathcal{F}_{p}(U,Y)=(\tr ( V^{\dagger} V)^{p/2} ))^{1/p},  V^{I}= Y^{I}(t) M_{I},  \quad p \textrm{\quad is\, an\, integer}\,\footnotemark.
\end{split}
\end{align}
 \footnotetext{These are formally known as ``Schatten norms" and first considered in \cite{Hackl:2018ptj} but explored in detail in \cite{recentmyers}.}
Here $p_{I}$ are some weights which at this moment are arbitrary.  We observe that $\mathcal{F}_{\kappa=1}$ is directly related to the number of gates that has been used to achieve the target state (at least upto a certain tolerance  $|| |\psi^{T}\rangle-U|\psi^{R}\rangle|| < \epsilon,$ where $\epsilon$ is a small and  adjustable parameter). On the other hand $\mathcal{F}_{2}$ with $p_I=1$ for all $I$ is basically a distance function on a given manifold.  These calculations have been generalized for free fermions in \cite{Khan:2018rzm, Hackl:2018ptj}. Interesting connections were found with holographic proposals in spite of the fact that these were free theories. \par  

It was suggested in \cite{Swingle} that tensor network approach maybe useful to derive qualitative features of holography, especially to understand the notion of emergent geometry from the field theory.  One such useful tool is cMERA (``{\it Continuous Multiscale Entanglement Renormalization Ansatz"}) \cite{cmera} which provide us with several interesting features of holography \cite{cmera1,cmera2,cmera3}. In \cite{Chapman1}, the authors have computed complexity for the cMERA circuit for ground state of  free scalar theory. Basically for the free scalar theory the  ground state wavefunction furnished by cMERA can be parametrized  as an $SU(1,1)$  coherent state. Then complexity  was computed by first defining a Fubini-Study metric  for $SU(1,1)$ manifold and computing the length of the geodesic with a suitable boundary condition. Further motivated by the tensor network representation  of the partition function an alternative  method of computing complexity for conformal field theory has been proposed in \cite{Cap1,Cap2} based on an {\it ``optimization"} procedure which basically determines how to represent most efficiently a partition function for conformal field theories. Then the complexity can be computed by evaluating simply the Liouville action  \cite{Cap1,Cap2, bartek, cap4}. This method has been generalized to include perturbations for 2 dimensional spacetimes \cite{Cap3}. These methods give the same leading divergent term for the complexity \cite{jm}.

Our goal in this paper is to consider interacting scalar field theories. A primary motivation is to try to establish a connection with the renormalization group perspective. Our progress in this front will be modest. We will compute circuit complexity in a variety of interesting interacting scalar QFTs to leading order in perturbation. We will generalize the approach of \cite{jm} to the interacting case. In the process of doing so, we will encounter several subtleties. For starters, since our approach will be based on the group $GL(N,R)$ we will not be able to use a purely Gaussian reference state for reasons we will explain. Rather we will forced to start with a slightly non-Gaussian reference state. This will lead to interesting technical complications. Namely we will find that we will be forced to make the cost functional dependent on the perturbative coupling so that we can smoothly interpolate to the free theory.

Taking into account these complications, we will then turn to evaluating circuit complexity in various dimensions including fractional dimensions to make a connection with the Wilson-Fisher fixed point in the epsilon expansion. We find that while the free theory depends linearly on the spatial volume, the interacting part shows a fractional volume dependence. Next we find that as dimensionality increases, the circuit depth also increases in the presence of the interaction (for positive coupling). In the RG paradigm, we know that the Gaussian fixed point is stable for $d>4$ while the Wilson-Fisher fixed point is stable for $d<4$. From the perspective of a potential quantum computer, we then find that since the circuit depth corresponding to turning on a coupling increases (and in fact diverges worse than the free theory eventually) with increasing dimensionality, it will be harder to perform the corresponding computation. Therefore, it appears to us that circuit complexity can be used as a diagnostic to analyse RG flows. Eventually, one can then hope that there could exist a monotonicity property much like the $c$-theorems in quantum field theory \cite{cthm}.

The fact that there must be an interesting connection between the renormalization group and complexity is not unexpected. In the context of the kind of calculations that were initiated in \cite{jm}, one can easily see this as follows. If we turn on a perturbative coupling, then clearly the complexity answer will get modified by this coupling. In the context of renormalized perturbation theory, the coupling is a function of the RG scale. As a result we can consider writing down a differential equation for the circuit complexity in terms of this scale. Since we are considering first order perturbation theory, the differential equation will relate circuit complexity to the beta function of the theory as well as the flow equation for the mass parameter. While this is unsurprising, what is important to know is the nature of this relation--in particular, is the perturbative approach to circuit complexity well defined in any dimension? Can we identify fixed points and establish if a fixed point is stable or unstable by considering circuit complexity? For instance intuitively we may expect that if a fixed point was stable, then moving away from this fixed point would increase the complexity, while if it was unstable then the reverse would happen. We will attempt to take some modest steps in these directions. A potentially useful spin-off in our investigation is that we will come up with a general analytic method to perform the required lattice sums. This will enable us to consider even fractional dimensions. This method is outlined in Appendix~(C) and will be heavily used in the paper.

The plan of the paper is as follows: In Section~(2) we discuss the $\lambda\,\phi^4$ on lattice as coupled Harmonic oscillators and solve its ground state wavefunction. Then we detail the complexity calculation for the the two oscillator case by generalizing the arguments of \cite{jm}. We discuss in detail all the subtleties regarding our approach and the construction of the circuit. In Section~(3) we generalize this for the arbitrary $N$ oscillator case  for  arbitrary spacetime dimensions $d$ and take the continuous limit  of the expression for the complexity. Then we derive flow equations for the complexity and study its implications. In Section~(4) we generalize all these for theories with  arbitrary number ($\mathcal{N}$) of scalar fields. In Section~(5) we discuss  briefly computation of complexity of $\phi^4$ theory using a different set of gates. Then we end with a summary and a list of some of interesting future problems. All the supplementary materials which we have deemed  useful for the reader have been placed in the appendices. 

\section{ Circuit complexity with $\phi^4$ interaction--the 2-oscillator case}
In this paper we will consider a massive scalar field theory with a $\hat\lambda\,  \phi^4$ interaction term. We will follow the notation in \cite{jm} to facilitate an easy comparison. The Hamiltonian for the theory is, 
\be
\mathcal{H}=\frac{1}{2}\int d^{d-1}x \Big[\pi(x)^2+( \nabla \phi(x))^2+m^2\phi(x)^2+\frac{\hat \lambda}{12}\phi(x)^4\Big]
\ee
where $d$ is the spacetime dimensions. We will assume the coupling  $\hat\lambda \ll 1$ so that we can work in a perturbative framework. $\hat \lambda$ is dimensionful having mass dimension $4-d$. Next we discretize this theory on a $d-1$ dimensional lattice. After discretization the Hamiltonian takes the following form, 
\be
\mathcal{H}=\frac{1}{2}\sum_{\vec n} \Big\{\frac{\pi(\vec n)^2}{\delta^{d-1}}+\delta^{d-1}\Big[\frac{1}{\delta^2}\sum_{i}(\phi(\vec n)-\phi(\vec n-\hat x_i))^2+m^2\phi(\vec n)^2+\frac{\hat \lambda}{12}\phi(\vec n)^4\Big]\Big\}.
\ee
$\vec n$ denotes the location of the points on the lattice.
Then using,  
\begin{align}
\begin{split} \label{eq3.3}
X(\vec n)=\delta^{d/2} \phi(\vec n), \, P(\vec n)=\pi(\vec n)/\delta^{d/2},\, M=\frac{1}{\delta}, \omega=m, \, \Omega=\frac{1}{\delta},\,  \lambda=\frac{\hat \lambda }{24}\delta^{-d}.
\end{split}
\end{align}
we arrive at the following\footnote{Note that $[\lambda]=4$ and $[\hat\lambda]=4-d$ in our notation. While introducing $\Omega$ seems redundant, it will facilitate a comparision with the coupled harmonic oscillator case as in \cite{jm} and we will continue using it.},
\begin{align}
\begin{split}
\mathcal{H}=\sum_{\vec n}\Big\{\frac{P(\vec n)^2}{2 M}+\frac{1}{2}M\Big[\omega^2 X(\vec n)^2+\Omega^2 \sum_{i} (X(\vec n)-X(\vec n-\hat x_{i}))^2+2 \lambda\, X(\vec n)^4\Big]\Big\}.
\end{split}
\end{align}
We will focus on evaluating complexity for the ground state of this Hamiltonian. It is evident that this system is nothing but coupled anharmonic oscillators.  For simplicity, first we focus on two coupled  oscillators. The Hamiltonian is:
\be \label{eq3.5}
\mathcal{H}=\frac{1}{2}\Big[p_{1}^2+p_{2}^2+\omega^2(x_1^2+x_2^2)+\Omega^2 (x_1-x_2)^2+2 \lambda\, (x_1^4+x_2^4)\Big].
\ee
We have also set $M=1$ here to make the analysis of the harmonic oscillator case more convenient--this will not affect the final answers which have to be dimensionally correct. Eigenstates of this Hamiltonian can be easily solved in normal mode coordinates. 
\begin{align}
\begin{split}\label{eq3.6}
&\tilde x_{0}=\frac{1}{\sqrt{2}}(x_1+ x_2), \quad \tilde x_{1}=\frac{1}{\sqrt{2}}(x_1-x_2),\\& \tilde p_{0}=\frac{1}{\sqrt{2}}(p_1+ p_2),\quad \tilde p_{1}=\frac{1}{\sqrt{2}}(p_1-p_2),\\& \tilde \omega^2_{0}=\omega^2, \quad\tilde \omega^2_{1}=\omega^2+2 \Omega^2.
\end{split}
\end{align}
We define
\begin{align}
\begin{split}\label{eq3.12}
\langle \tilde x_0,\tilde x_1|\psi^{0}(n_1,n_2)\rangle=&\psi_{n_1, n_2}^{0}(\tilde x_{0},\tilde x_{1})\\&=\frac{1}{\sqrt{2^{n_1+n_2}n_1!n_2!}}\frac{(\tilde \omega_{0}\tilde \omega_{1})^{1/4}}{\sqrt{\pi}} e^{-\frac{1}{2}\tilde \omega_{0}\tilde x_{0}^2-\frac{1}{2}\tilde \omega_{1}\tilde x_{1}^2}H_{n_1}(\sqrt{\tilde \omega_{0}}\tilde x_{0})H_{n_2}(\sqrt{\tilde \omega_{1}}\tilde x_{1})\,,
\end{split}
\end{align}
where, $H_{n}(x)'s$ are the Hermite polynomials with  $H_{0}(x)=1$.  The expression for the ground state eigenfunction to first order in $\lambda$ can be written as
\be
\psi_{n_1,n_2}(\tilde x_0,\tilde x_1)=\psi^{0}_{0,0}(\tilde x_0,\tilde x_1)+\lambda\, \psi_{0,0}^{1}(\tilde x_{0},\tilde x_{1})\,,
\ee
with
\be
\psi^{0}_{0,0}(\tilde x_0,\tilde x_1)=\frac{(\tilde \omega_{0}\tilde \omega_{1})^{1/4}}{\sqrt{\pi}} e^{-\frac{1}{2}\tilde \omega_{0}\tilde x_{0}^2-\frac{1}{2}\tilde \omega_{1}\tilde x_{1}^2},
\ee
and
\begin{align}
\begin{split}
\psi_{0,0}^{1}(\tilde x_{0},\tilde x_{0})&=-\frac{3(\tilde \omega_{0}+\tilde \omega_{1})}{4\sqrt{2}\, \tilde \omega_{0}^3\tilde  \omega_{1}}\psi_{2,0}^{0}(\tilde x_0,\tilde x_1)-\frac{3(\tilde \omega_{0}+\tilde \omega_{1})}{4\sqrt{2}\, \tilde \omega_{0} \tilde \omega_{1}^3}\psi_{0,2}^{0}(\tilde x_0,\tilde x_1)\\&-\frac{3}{4\tilde \omega_{0}\tilde \omega_{1}(\tilde \omega_{0}+\tilde \omega_{1})}\psi^{0}_{2,2}(\tilde x_0,\tilde x_1)- \frac{\sqrt{3}}{8\sqrt{2}\,\tilde \omega_{0}^3}\psi^{0}_{4,0}(\tilde x_0,\tilde x_1)- \frac{\sqrt{3}}{8\sqrt{2}\,\tilde \omega_{1}^3}\psi^{0}_{0,4}(\tilde x_0,\tilde x_1).
\end{split}
\end{align}
For later convenience we will use 
\be \label{eq3.17}
\psi_{0,0}(\tilde x_1,\tilde x_{2})\approx \frac{(\tilde \omega_{0}\tilde \omega_{1})^{1/4}}{\sqrt{\pi}}\exp(a_0)\exp\Big[-\frac{1}{2}\Big(a_1 \tilde x_0^2+a_2 \tilde x_1^2+ a_3 \tilde x_0^4+a_4\tilde x_1^4+ a_5 \tilde x_0^2 \tilde x_1^2\Big)\Big]
\ee
where,
\begin{align}
\begin{split} \label{eq3.18}
&a_0=\frac{3\lambda}{8}\Big(\frac{3}{4\tilde \omega_0^3}+\frac{3}{4\tilde \omega_{1}^3}+\frac{\tilde \omega_0\tilde \omega_1+\tilde \omega_0^2+\tilde \omega_1^2}{\tilde \omega_0^2\tilde \omega_1^2(\tilde \omega_0+\tilde \omega_1)}\Big),
\quad a_1=\tilde \omega_0+\frac{1}{\tilde \omega_0}\Big(3 a_3+\frac{a_5}{2}\Big),\\&
a_2=\tilde \omega_1+\frac{1}{\tilde \omega_1}\Big(3 a_4+\frac{a_5}{2}\Big),\quad
a_3=\frac{\lambda }{4\tilde \omega _0},\quad
a_4=\frac{\lambda }{4 \tilde \omega _1},\quad
a_5=\frac{3 \lambda }{ \left(\tilde \omega _1+\tilde \omega _0\right)},\end{split}
\end{align}
where it is understood that the expression in (\ref{eq3.17}) can be trusted only upto linear order in $\lambda.$ We will also use an approximate wavefunction \footnote{This kind of approximate Gaussian wavefunction for interacting quantum fields has  also been used in some version of cMERA \cite{Cotler1}.}
\be \label{approx}
\widetilde{\psi_{0,0}}(\tilde x_1,\tilde x_{2})\approx \frac{(\tilde \omega_{0}\tilde \omega_{1})^{1/4}}{\sqrt{\pi}}\exp(a_0)\exp\Big[-\frac{1}{2}\Big(a_1 \tilde x_0^2+a_2 \tilde x_1^2\Big)\Big].
\ee
Defining the fidelity as 
\be
F(1,2)=1-\frac{|\langle 1|2\rangle|^2}{\langle 1|1\rangle \langle 2|2\rangle}\,,
\ee
we find that 
\be 
F(\psi,\widetilde\psi)=\frac{3\, \lambda }{32}  \left(\frac{1}{\tilde \omega_{0}^3}+\frac{4}{\tilde \omega_{0}\tilde \omega_{1}(\tilde \omega_{0}+\tilde \omega_{1})}+\frac{1}{\tilde \omega_{1}^3}\right)\,.
\ee 
 \subsection*{Circuit complexity}
 We would like to evaluate the circuit complexity for this ground state wavefunction starting from a reference state.  
 This needs several assumptions on our part. We will need to specify the reference state and available gates. 
We begin by writing the wavefunction in the following form:
 \begin{align}
\begin{split} \label{eq3.19}
\psi^{s}( \tilde x_0, \tilde x_1)=\mathcal{N}^{s}\,\exp \Big[-\frac{1}{2}(v_{a}. A(s)_{a\,b}.v_{b})\Big].
\end{split}
\end{align}
$\mathcal{N}^{s}$ is a normalization factor. $s$ is a running parameter and parametrizes the space of circuits. For $s=1$ this coincides with the target state  (\ref{eq3.17}) with $\mathcal{N}^{s=1}=\frac{(\tilde \omega_{0}\tilde \omega_{1})^{1/4}}{\sqrt{\pi}}\exp(a_0)$, while $s=0$ will be the reference state. Here the idea is to write the exponent of the wavefunction as a matrix $A(s)$ conjugated by a basis vector $\vec{v}.$ If the state is just a Gaussian state then $\vec{v}=\{\tilde x_0, \tilde x_1\}$ only. But for our case the states are not Gaussian so we have to extend the definition of this basis vector $\vec{v}.$ Below we discuss this explicitly for our setup.

A desirable property of the reference state is that it should not contain any entanglement in the original coordinates. Also we have to keep in mind that in order to represent it in  the form (\ref{eq3.19}), we have to allow non-linear terms ($x_1^2$ and $x_2^2$ )   in the basis vector. Keeping in mind these two facts, we choose the following, 
\begin{align}
\begin{split} \label{refstate}
\psi^{s=0}(x_1,x_2)=\mathcal{N}^{s=0}\exp\Big[-\frac{\tilde \omega_{ref}}{2}(x_1^2+x_2^2+\lambda_0(x_1^4+x_2^4))\Big].
\end{split}
\end{align}
Here $\lambda_0$ is some parameter that we will fix later on. It parametrizes a non-Gaussianity in the reference wavefunction.
Now going to the normal coordinates we get,
\begin{align}  \label{eq3.21a2}
        \begin{split}
        \psi^{s=0}(\tilde x_0,\tilde x_1)=\mathcal{N}^{s=0}\exp\Big(-\frac{\tilde \omega_{ref}}{2}\Big[\tilde x_0^2+\tilde x_1^2+\frac{\lambda_{0}}{2}(\tilde x_0^4+ \tilde x_1^4+6 \tilde x_0^2\tilde x_1^2)\Big]\Big).
        \end{split}
        \end{align}
Now we rewrite this state in the form (\ref{eq3.19}). We also want to make sure that the matrix $A$ is non singular. We have to  choose the following basis (this is one of the choices and the minimal one as far as we can see)

\be \label{basis}
 \vec{v}= \{\tilde x_{0},\tilde x_{1},\tilde x_0\,\tilde x_1, \tilde x_{0}^2,\tilde x_{1}^2\}.
 \ee In this basis,
 \be \label{eq3.21a1}
        A(s=0)=\left(
\begin{array}{ccccc}
 \tilde \omega_{ref} & 0 & 0 & 0 & 0 \\
 0 & \tilde \omega_{ref} & 0 & 0 & 0 \\
 0 & 0 & b \lambda_{0} \tilde \omega_{ref} & 0 & 0 \\
 0 & 0 & 0 & \frac{\lambda_{0} \tilde \omega_{ref}}{2} & \frac{1}{2} (3-b) \lambda_{0} \tilde \omega_{ref} \\
 0 & 0 & 0 & \frac{1}{2} (3-b) \lambda_{0} \tilde \omega_{ref} & \frac{\lambda_{0} \tilde \omega_{ref}}{2} \\
\end{array}
\right).
        \ee
$b$ is arbitrary. Now given the basis  vector $\vec{v}$ in (\ref{basis}) we can easily verify that both for the reference and the target state we can write the  exponents  in the form $v_{a} A_{a\, b} v_{b}$ as advocated in (\ref{eq3.19}). Also we can see that to write the exponents in this form it is \textit{absolutely} necessary to include quadratic terms in the basis (\ref{basis}). \par
Next  we want to make the determinant of this matrix $A$  positive which needs $2 < b <4.$ Now further to make our analysis simple we choose $$b=3$$ to kill  the off-diagonal components in the reference state.   Now the  target state in this basis will look like,

\be \label{eq3.21}
        A(s=1)=\left(
\begin{array}{ccccc}
 a_1& 0 & 0 & 0 & 0 \\
 0 & a_2 & 0 & 0 & 0 \\
 0 & 0 & \tilde b\, a_5 & 0 & 0 \\
 0 & 0 & 0 & a_3 & \frac{1}{2} (1-\tilde b)a_5 \\
 0 & 0 & 0 & \frac{1}{2} (1-\tilde b) a_5 & a_4\\
\end{array}
\right).
 \ee
The coefficients are given in (\ref{eq3.18}). Also we  restrict $\tilde b$ such that the determinant of this matrix is positive definite. As $\tilde \omega_1>\tilde \omega_0,$ we have $a_3>a_4>0.$ Also $a_5>0.$  These conditions together with fact that the eigenvalues are positive lead to,
\be \label{renage}
1-\frac{1}{6} \sqrt{\frac{(\tilde \omega_0+\tilde \omega_0)^2}{\tilde \omega_0\tilde \omega_1}}<\tilde b<1+\frac{1}{6} \sqrt{\frac{(\tilde \omega_0+\tilde \omega_0)^2}{\tilde \omega_0\tilde \omega_1}}.
\ee
The upper limit is always positive but depending on the values of $\tilde \omega_0,\tilde \omega_0,$ $1-\frac{1}{6} \sqrt{\frac{(\tilde \omega_0+\tilde \omega_0)^2}{\tilde \omega_0\tilde \omega_1}}$ can be both positive or negative. But we have to make sure that $\tilde b\, a_5$ is also  positive to make the determinant of $A(s=1)$ positive. So we restrict ourselves to, 
\be \label{range1}
0<\tilde  b< 1+\frac{1}{6} \sqrt{\frac{(\tilde \omega_0+\tilde \omega_0)^2}{\tilde \omega_0\tilde \omega_1}}.
\ee
Now  starting from this reference state the target state can be achieved via unitary evolution
\be\label{eq3.23}
\psi^{s=1}(\tilde x_{0},\tilde x_1)=U(s=1) \psi^{s=0}(\tilde x_0,\tilde x_1).
\ee
The unitary takes the following form,
\be\label{eq3.24}
U(s)={\overleftarrow{\mathcal{P}}}\exp\Big(\int_{0}^{s} ds\, Y^I(s) \, \mathcal{O}_{I}\Big).
\ee
As explained before we have to act the reference state by the set of the operators $\mathcal{O}_{I}'s$ in a particular sequence. Note that, the $Y^{I}(s)'s$ depend on the path, i.e.,  on the particular sequence in which $O_{I}$'s are acting on the reference state. Our target is to find the shortest possible path such that we will achieve minimum complexity. To do so, we try to attach a geometrical interpretation to this process \cite{jm, Nielsen1, Nielsen2, Nielsen3}. Now looking at the structure of the matrix $A$ in (\ref{eq3.21}) we can  consider $U(s)$ to be an element of  $GL(5,R)$  with positive determinant. Then we can write $U(s)$ in the following way

\be
U(s)=\overleftarrow{\mathcal{P}}\exp\Big(\int_{0}^{s} Y^{I}(s)M_{I} ds\Big),
\ee
where, $(M_{I})_{jk}'s$ are $GL(5, R)$ generators satisfying,
\be
Tr[M_{I} M_{J}]=2 \delta_{IJ}.
\ee
$I, J$ runs from $1$ to $25.$
   Also we have \footnote{Now $U$ when acting on $A$ via eq.(\ref{eq3.31}) is not unitary! Rather what happens is that this $U$ can be mapped to a unitary operator which then acts on the wavefunction as we will discuss. This is a notational issue in \cite{jm} that we will assume does not create too much confusion.}, 
\be \label{eq3.31}
A(s=1)=U(s=1) \,A(s=0) \,U^{T}(s=1).
\ee
From this we get,
\be
Y^{I}M_{I}=\partial_{s} U(s) U(s)^{-1}.
\ee
Hence,
\be \label{eq3.33}
Y^{I}=\frac{1}{Tr(M^I (M^{I})^{T})}Tr(\partial_s U(s) U^{-1} (M^{I})^{T}).
\ee
Next the infinitesimal distance i.e., metric in the parameter space  defined by $Y^{I}$'s can be written  as,
\begin{align}
\begin{split} \label{eq3.34}
ds^2=&\,\, G_{IJ} dY^{I} dY^{J},\\& =\,G_{IJ}\Big(\frac{1}{Tr( M^{I} ( M^{I})^{T})}Tr(d_s U(s) U^{-1} ( M^{I})^{T})\Big)\Big(\frac{1}{Tr( M^{J} ( M^{J})^{T})}Tr(d_s U(s) U^{-1} (M^{J})^{T})\Big).
\end{split}
\end{align}
At this stage we have various choices for $G_{IJ}.$ First, for simplicity we set $G_{IJ}$ as $25 $ by $25$ identity matrix.  Now $U(s)$ is an element of $GL(5, R).$ To do further calculations we have to chose a suitable parametrization for $U(s)$. In general any element ($g$) of $GL(5, R)$ can be written in terms of product of an orthogonal  matrix, a diagonal matrix and an upper triangular matrix ($G=K A N$), which is known as Iwasawa decomposition. 
But after closely inspecting our target state (\ref{eq3.21}),  we can further simplify our choice of parametrization. We can easily infer that our target state is of a block diagonal form. This motivates us to parametrize  $U(s)$ in the following way,  
    \begin{align}
\begin{split}
U(s)= \left(
\begin{array}{ccccc}
x_0-x_3& x_2-x_1& 0 & 0 &0 \\
x_2-x_1&  x_0+x_3 & 0 & 0 &0\\
0&0&\exp[y_2(s)]&0&0\\
 0 & 0 & 0&\tilde x_{0}-\tilde x_{3} & \tilde x_{2}-\tilde x_{1} \\
 0&0&0&\tilde x_{1}+\tilde x_{2} & \tilde x_{0}+\tilde x_{3} \\
\end{array}
\right).
\end{split}
\end{align}
We have decomposed $U(s)$ in terms of $ R^3\times GL(2, R)$ blocks. We could have allowed for  off-diagonal elements. But it is evident that at the end of the evolution those terms have to vanish as the final state is also in the block diagonal form, so allowing those off-diagonal terms will only increase the path length and hence the complexity.  Now $GL(2, R)$ can be written as $R \times SL(2, R).$ We also note that the first block in (\ref{eq3.21}) is diagonal. It is expected that in the normal mode coordinates the quadratic part of the target state (\ref{eq3.17}) is always diagonal. As argued previously\cite{jm}, this induces a flat metric.  Keeping this is in mind we can set, $x_1=x_2=0.$ 
For the rest of the components we choose the following parametrization,
\begin{align}
\begin{split}\label{eq3.36}
&x_0=\exp(y_1(s))\cosh(\rho_1(s)),x_3=\exp(y_1(s))\sinh(\rho_1(s)),\\&
\tilde x_0=\exp(y_3(s))\cos(\tau_3(s))\cosh(\rho_3(s)),\tilde x_1=\exp( y_3(s))\sin(\tau_3(s))\cosh(\rho_3(s)),\\&
\tilde x_2=\exp(y_3(s))\cos(\theta_3(s))\sinh(\rho_3(s)),\tilde x_3=\exp(y_3(s))\sin(\theta_3(s))\sinh(\rho_3(s)).
\end{split}
\end{align}
Then the metric in (\ref{eq3.34}) becomes, 
\begin{align}
\begin{split}\label{eq2.37}
ds^2=&2\Big( dy_1^2+dy_2^2+dy_3^2+d\rho_1^2+d \rho_3^2\\&+\cosh(2\rho_3)\sinh^2(\rho_3) d\theta_3^2+\cosh(2 \rho_3)\cosh^2(\rho_3) d\tau_3^2-\sinh^2(2\rho_3) d\theta_3 d\tau_3\Big).
\end{split}
\end{align}
A suitable functional for this case is 
\be \label{anew1}
\mathcal{F}_{2}(U,Y^{I})=\sqrt{\sum_{I}\Big|Y^{I}(s)\Big|^2}.\ee
Using (\ref{eq2.37}) we find
 \begin{align}
\begin{split}\label{eq3.38}
\mathcal{D}(U)=&\int_{0}^{1} \mathcal{F}_2(U,Y^{I}(s)) ds=\int_{0}^{1} ds\,\sqrt{g_{ij}\dot x^{i}\dot x^{j}},\\&
=\int_0^1ds\,\Big(2\Big[\Big( \frac{d y_1}{ds}\Big)^2+\Big( \frac{d \rho_1}{ds}\Big)^2+\Big(\frac{d y_2}{ds}\Big)^2+\Big( \frac{d y_3}{ds}\Big)^2+\Big( \frac{d \rho_3}{ds}\Big)^2\\&-\sinh^2(2\,\rho_3)\Big( \frac{d \theta_3}{ds}\Big)\Big( \frac{d \tau_3}{ds}\Big)+ \cosh(2\,\rho_3)\Big\{\sinh^2(\rho_3) \Big( \frac{d \theta_3}{ds}\Big)^2+\cosh^2(\rho_3)\Big( \frac{d\tau_3}{ds}\Big)^2\Big\}\Big]\Big).
\end{split}
\end{align}

 The problem of finding the shortest path will then be mapped to the problem of finding geodesic in $GL(5, R)$ group manifold. So we have to find the geodesic coming from extremizing (\ref{eq3.38}) and evaluate (\ref{eq3.38}) on this geodesic.  The boundary conditions from (\ref{eq3.31}) are
\be \label{eq3.39}
\rho_1(0)=\rho_3(0)=y_1(0)= y_2(0)=y_3(0)=0,
\ee 
and 
\begin{align}
\begin{split}\label{eq3.40}
&2 (y_1(1)-\rho_1(1))=\log[\frac{a_1}{\tilde \omega_{ref}}]=\log\Big[\frac{3 \lambda  \left(\tilde \omega _1+3\tilde \omega _0\right)}{4\, \tilde \omega_{ref} \tilde \omega _0^2 \left(\tilde \omega _1+\tilde \omega _0\right)}+\frac{\tilde \omega _0}{\tilde \omega_{ref}}\Big],\\&
2(y_1(1)+\rho_1(1))=\log[\frac{a_2}{\tilde \omega_{ref}}]= \log\Big[\frac{3 \lambda  \left(3\tilde \omega _1+\tilde \omega _0\right)}{4\, \tilde \omega_{ref} \tilde  \omega _1^2 \left(\tilde \omega _1+\tilde \omega _0\right)}+\frac{\tilde \omega _1}{\tilde \omega_{ref}}\Big],\\&
2 y_3(1)=\log\Big[2\frac{\sqrt{a_3a_4-\frac{(1-\tilde b)^2a_5^2}{4}}}{\tilde \omega_{ref}\lambda_{0}}\Big]= \log\Big[\frac{2\lambda\, \sqrt{\frac{1 }{16\tilde \omega _1\tilde \omega _0}-\frac{9 (1-\tilde b)^2 }{4 \left(\tilde \omega _0+\tilde \omega _1\right)^2}}}{\tilde \omega_{ref}\lambda_0}\Big],\\&
2 \rho_3(1)=\cosh^{-1}\Big[\frac{a_3+a_4}{2\sqrt{a_3 a_4-\frac{(1-\tilde b)^2a_5^2}{4}}}\Big]=\cosh^{-1}\Big[\frac{\tilde \omega_0+\tilde\omega_{1}}{8\tilde \omega_{0}\tilde \omega_{1} \sqrt{\frac{1 }{16\tilde \omega _0\tilde \omega _1}-\frac{9  (1-\tilde b)^2 }{4 \left(\tilde \omega _0+\tilde \omega _1\right)^2}}}\Big],\\&
2y_2(1)=\log\Big[\frac{\tilde b\, a_5}{3\,\lambda_0\, \tilde\omega_{ref}}\Big]=\log\Big[\frac{\tilde b\,\lambda}{\lambda_0\tilde \omega_{ref}(\tilde \omega_0+\tilde \omega_1)}\Big].
\end{split}
\end{align}
 Now we proceed to solve the geodesic equations. These equations can be solved by first finding the conserved charges. As our metric is nothing but the tensor product of $R^4$ and $SL(2, R)$  matrices, we essentially use the results of \cite{jm}. Then we get after using (\ref{eq3.39}) and  (\ref{eq3.40}),
\be
y_1(s)=y_1(1) s, \, \rho_1(s)=\rho(1)\,s, \, y_{3}(s)= y_3(1) s, \, \rho_{3}(s)=\rho(1) s.\ee
Also using similar arguments as \cite{jm} we can set 
\begin{align}
\begin{split}
 \tau_3(s)=0, \theta_3(s)=\theta_0.
\end{split}
\end{align}
$\theta_0$ is just a constant independent of $s.$ We have trivially,
\be
y_2(s)=y_2(1) s.
\ee
  Collecting all these together finally our complexity functional evaluates to,
\begin{align}
\begin{split}
\mathcal{D}(U)=
&\Big(\sqrt{ 2(y_1(1)^2+\rho_1(1)^2+y_2(1)^2+ y_3(1)^2+\rho_3(1)^2)}\Big)
=\frac{1}{2}\Big(\sqrt{d}\Big)+\frac{\lambda\, c}{2\sqrt{d}}+\mathcal{O}(\lambda^2),
\end{split}
\end{align}
where, \begin{align}
\begin{split} \label{eq3.47}
&c=\Big(\frac{3  \left(\tilde \omega _1+3\tilde \omega _0\right)\log[\frac{\tilde \omega _0}{\tilde \omega_{ref}}]}{4\,\tilde \omega _0^3 \left(\tilde \omega _1+\tilde \omega _0\right)}+\frac{3  \left(\tilde \omega _0+3\tilde \omega _1\right)\log[\frac{\tilde \omega _1}{\tilde \omega_{ref}}]}{4\,\tilde \omega _1^3 \left(\tilde \omega _1+\tilde \omega _0\right)}\Big),\\&d=\Big(\log^2\Big[\frac{\tilde \omega _0}{\tilde \omega_{ref}}\Big]+\log^2\Big[\frac{\tilde \omega _1}{\tilde \omega_{ref}}\Big]\Big)+2\,\cosh^{-1}\Big[\frac{\tilde \omega_0+\tilde\omega_{1}}{8\tilde \omega_{0}\tilde \omega_{1} \sqrt{\frac{1 }{16\tilde \omega _0\tilde \omega _1}-\frac{9 (1-\tilde b)^2  }{4 \left(\tilde \omega _0+\tilde \omega _1\right)^2}}}\Big]\\&~~~~~~~+2\, \log\Big[\frac{2\lambda\, \sqrt{\frac{1 }{16\tilde \omega _1\tilde \omega _0}-\frac{9 (1-\tilde b)^2 }{4 \left(\tilde \omega _0+\tilde \omega _1\right)^2}}}{\tilde \omega_{ref}\lambda_0}\Big]+2\log\Big[\frac{\tilde b\,\lambda}{\lambda_0\tilde \omega_{ref}(\tilde \omega_0+\tilde \omega_1)}\Big].
\end{split}
\end{align}
We can evaluate $U(s)=\exp\Big(\tilde M s\Big) $ on this solution which will give us the optimal circuit. From there we can also identify the   unitary operators which are  the building blocks of this optimal circuit. 
 Unitaries that we are considering are of the  form $\exp(i\, O),$ where $O$ is Hermitian. Then  by acting $i\, O$ on the basis vector $\vec{v}$ as defined in (\ref{basis}) we can identify these $i\, O's$ with  various matrices  $M's$  and given those matrices we can show that $U(s)=\exp\Big(\tilde M s\Big) $ where, $\tilde M$ is given by the following linear combination,
\begin{align}
\begin{split} \label{lincomb}
\tilde M=\alpha  M_{11}+\beta  M_{22}+\gamma  M_{33}+\delta  M_{44}+\zeta  M_{55}+\tau  M_{66} +\kappa  M_{77}+\mu M_{88}
\end{split}
\end{align}
by suitably choosing the parameters $\alpha,\beta,\gamma,\delta,\zeta,\tau ,\kappa$ and $\mu.$ Below we identify the $M$'s with the corresponding $iO$'s and give the details in the Appendix~(A). 
\begin{align}\label{cgates}
\begin{split}
 &\frac{ i\,(\tilde x_0\tilde p_0+\tilde p_0\,\tilde x_0)}{2}=i\,(\tilde x_0\,\tilde p_0-\frac{i}{2}),\quad  i\, \tilde x_0\,\tilde p_0\rightarrow M_{11},
\frac{ i\,(\tilde x_1\tilde p_1+\tilde p_1\,\tilde x_1)}{2}=i\,(\tilde x_1\,\tilde p_1-\frac{i}{2}),\quad  i\, \tilde x_1\,\tilde p_1\rightarrow M_{22},\\&
\frac{i}{8}\, (\tilde x_1\tilde p_1+\tilde p_1\tilde x_1)(\tilde x_0\tilde p_0+\tilde p_0\tilde x_0)(\tilde x_0\tilde p_0+\tilde p_0\tilde x_0)\\&=\frac{i}{8}(8 \tilde x_1\tilde p_1(\tilde x_0 \tilde p_0)^2-8\, i\, \tilde x_1\tilde p_1\tilde x_0\tilde p_0-2\tilde x_1\tilde p_1-4\, i\, (\tilde x_0\tilde p_0)^2-4 \tilde x_0\tilde p_0+i ),\\& \frac{i}{8}(8 \tilde x_1\tilde p_1(\tilde x_0 \tilde p_0)^2-8\, i\, \tilde x_1\tilde p_1\tilde x_0\tilde p_0-2\tilde x_1\tilde p_1-4\, i\, (\tilde x_0\tilde p_0)^2-4 \tilde x_0\tilde p_0 ) \rightarrow M_{33},\\&
\frac{i}{8}(8 \tilde x_0\tilde p_0(\tilde x_1 \tilde p_1)^2-8\, i\, \tilde x_0\tilde p_0\tilde x_1\tilde p_1-2\tilde x_0\tilde p_0-4\, i\, (\tilde x_1\tilde p_1)^2-4 \tilde x_1\tilde p_1 ) \rightarrow M_{44},\\&
\frac{i}{8}\, (\tilde x_0\tilde p_0+\tilde p_0\tilde x_0)(\tilde x_0\tilde p_1+\tilde x_1\tilde p_0)(\tilde x_0\tilde p_1+\tilde x_1\tilde p_0) \rightarrow M_{55}, \frac{i}{8}\, (\tilde x_1\tilde p_1+\tilde p_1\tilde x_1)(\tilde x_0\tilde p_1+\tilde x_1\tilde p_0)(\tilde x_0\tilde p_1+\tilde x_1\tilde p_0) \rightarrow M_{66},\\&
\frac{i}{8}(8 \tilde x_0\tilde p_0(\tilde x_0 \tilde p_0)^2-8\, i\, \tilde x_0\tilde p_0\tilde x_0\tilde p_0-2\tilde x_0\tilde p_0-4\, i\, (\tilde x_0\tilde p_0)^2-4 \tilde x_0\tilde p_0 ) \rightarrow M_{77},\\
&\frac{i}{8}(8 \tilde x_1\tilde p_1(\tilde x_1 \tilde p_1)^2-8\, i\, \tilde x_1\tilde p_1\tilde x_1\tilde p_1-2\tilde x_1\tilde p_1-4\, i\, (\tilde x_1\tilde p_1)^2-4 \tilde x_1\tilde p_1 ) \rightarrow M_{88}.
\end{split}
\end{align}

Also we note that, it is evident from this analysis that we can recast the unitary $\tilde M$ by the linear combinations of only those operators which scale the coordinates. This is not surprising  given the fact, that we have chosen a non Gaussian reference state  which already contains quartic terms apart from the usual quadratic terms.  Hence we can reproduce the target state by simply scaling all these  terms. Now we end this section by making some comments.

 \subsubsection*{Comments}

\begin{itemize}
\item{First of all we note that unlike for the free theory\cite{jm}, we cannot just choose the reference state as a product of Gaussians. We have to allow $x_{0}^4$ and $x_{1}^4$ term in the reference state. If we try to set $\lambda_0 =0$, our boundary conditions will be ill-defined as evident from (\ref{eq3.40}), specially $\tilde y(1).$ This is also tied to the fact that we are setting up the problem using the machinery of $GL(N,R)$ group which required us to have the determinant of $A(s)$ to be non zero.  More specifically, given the choice of the basis (\ref{basis}), if we would chosen our reference state as Gaussian state then the matrix $A(s=0)$ will have zero determinant because of the absence of quartic terms in the wavefunction. Then the relation (\ref{eq3.31}) will not hold as the  conjugating by $U$ preserves the determinant of both the reference and target matrix $A.$ So this forces us to make $\lambda_0 \neq 0.$ Note that preparation of non-Gaussian states is a hard problem and only partial results exist in the quantum information literature \cite{nongauss}. }


\item  We can rewrite the  $\mathcal{F}_{2}$ given in (\ref{anew1}) in the following form:
 
  \be\label{evalm}
\mathcal{F}_{2}=\frac{1}{2}\Big(\sqrt{\log^2\Big(\frac{\lambda_1}{\tilde \omega_{ref}}\Big)+\log^2\Big(\frac{\lambda_2}{\tilde \omega_{ref}}\Big)+\Big[\log^2\Big(\frac{\lambda_3}{3\,\tilde \omega_{ref}\lambda_0}\Big)+\log^2\Big(\frac{\lambda_4}{\frac{\tilde \omega_{ref}\lambda_0}{2}}\Big)+\log^2\Big(\frac{\lambda_5}{\frac{\tilde \omega_{ref}\lambda_0}{2}}\Big)}\Big]\Big)\,.
\ee
 Here $\lambda_1, \lambda_2, \lambda_3, \lambda_4, \lambda_5$ are the eigenvalues of the target matrix $A(s=1)$. \par
 
 Since $A(s=1)$ and $A(s=0)$ are real symmetric matrices that commute, therefore they both can be expressed as diagonal matrices in a common basis with corresponding eigenvalues as diagonal entries. This diagonalization is brought about by the action of a unitary matrix $R$ via a similarity transformation given by $R.A(s=0/1).R^T$. More details of this can be found in \cite{jm}. It can also be shown that, under this transformation the metric remains invariant and hence the complexity. Then the complexity is simply given in terms of the eigenvalues of the matrix $A$ of the target state. This is expected because once both the reference and target matrices are diagonal in a common basis all that we need are scaling gates corresponding to this diagonal basis to take us from the  reference to the target state. This is because any non diagonal entries introduced by the entangling gates corresponding to the new diagonal basis, have to be nullified using more entangling gates before reaching the target state. Thus the use of entangling gates would only add more to the number of gates required and hence would not correspond to minimal complexity. Having said that it can easily be seen that the number of scaling gates say corresponding to the $i^{th}$ diagonal element required to reach from the reference to the target state is $\frac{1}{2}\log\Big(\frac{\lambda_i}{r_i}\Big)$ where $\lambda_i$ is the $i^{th}$ diagonal element of the target matrix and $r_i$ is the $i^{th}$ diagonal element of the reference matrix . Plugging this result into the complexity functional $\mathcal{F}_2$ results in equation \eqref{evalm}. These expressions can be easily generalized for arbitrary lattice size in arbitrary dimensions.

\item 
Now two of the eigenvalues $\lambda_1, \lambda_2$ coming from the quadratic part of the wavefunction, are of the form $a_{1,2}+\lambda\, b_{12}$. The other three eigenvalues $\lambda_{3,4,5}$ coming from the quartic part of the wavefunction are of the $\mathcal{O}(\lambda)$. Taking logarithm of these $\mathcal{O}(\lambda)$ eigenvalues gives $\log(\lambda)$ terms in \eqref{evalm}. This will make the $\lambda \rightarrow 0$ limit ill-defined. To avoid this problem we can  choose $\lambda_0$ to be proportional to $\lambda$ such that the $\lambda$ dependence inside the logarithm  cancels out. Further we would want that \eqref{evalm} can be expanded perturbatively in $\lambda.$
 
 \be
 \mathcal{F}_{2}= \mathcal{F}^{free}_{2}+\lambda\, \mathcal{F}^1_{2}+\mathcal{O}(\lambda^2).
 \ee
 
Since we are working in perturbation theory, we would expect to recover the free result by taking $\lambda\rightarrow 0$. The $\lambda \rightarrow 0$ limit of (\ref{eq3.47}) is subtle. We would have expect that in $\lambda\rightarrow 0$ limit we would recover the free theory result which is given by 
\be\label{eq3.48}
\mathcal{F}^{free}_{2}=\frac{1}{2}\Big(\log^2\Big[\frac{\tilde \omega _0}{\tilde \omega_{ref}}\Big]+\log^2\Big[\frac{\tilde \omega _1}{\tilde \omega_{ref}}\Big]\Big).\ee But this is not the case. There are actually two problems:
\begin{itemize}
\item{
The third term in the expression for $d$ in (\ref{eq3.47}) does not have any counterpart in the free theory,  but still it seems that in $\lambda \rightarrow 0$ limit it doesn't vanish.}

\item{ Also in the fourth term in the expression for $d$ in (\ref{eq3.47}) we get $\log(\lambda)$. This make the $\lambda\rightarrow 0$ limit ill-defined. }
\end{itemize}
 The second problem can be easily cured by making $\lambda_0 \propto \lambda.$ This will also give that, in $\lambda \rightarrow 0$ limit $A(s=0)$ will reduce to the product of Gaussians. 
 The first problem is harder to solve. We could envisage having a smooth $\lambda\rightarrow 0$ limit by choosing $G_{IJ}$ differently so that the appropriate components pertaining to the second block are proportional to $\lambda$ leading to 
 \be\label{Astuff}
\mathcal{F}_{2}=\frac{1}{2}\Big(\sqrt{\log^2\Big(\frac{\lambda_1}{\tilde \omega_{ref}}\Big)+\log^2\Big(\frac{\lambda_2}{\tilde \omega_{ref}}\Big)+\mathcal{A}\Big[\log^2\Big(\frac{\lambda_3}{3\,\tilde \omega_{ref}\lambda_0}\Big)+\log^2\Big(\frac{\lambda_4}{\frac{\tilde \omega_{ref}\lambda_0}{2}}\Big)+\log^2\Big(\frac{\lambda_5}{\frac{\tilde \omega_{ref}\lambda_0}{2}}\Big)}\Big]\Big)\,,
\ee
with ${\mathcal A}\propto \lambda$.
  However, this makes the procedure of determining the complexity somewhat ad hoc and introduces a plethora of possible circuits. Also note that apart from the gates corresponding to the generators $M_{11}, M_{22}$, the rest of the gates are complicated and hence must be difficult to ``manufacture.'' Hence it makes sense to consider a somewhat different problem where instead of the target state $\psi_{0,0}$ we will use the approximate target state $\widetilde \psi_{0,0}$ given in (\ref{approx}). This will solve the following problem: Given the gates corresponding to $M_{11}, M_{22}$ find the circuit complexity to go from a Gaussian reference state to the approximate ground state given in (\ref{approx}). 
This essentially  means that in (\ref{Astuff})
we drop the terms proportional to $\mathcal A$. 
This also essentially makes the geodesic problem identical to \cite{jm}. In what follows, when we compute complexity expressions from the first block or unambiguous block, this is what we are doing. We will then consider the terms proportional to $\mathcal A$ which means introducing more complicated gates as in eq.(\ref{cgates}). This will need us to choose suitable penalty factors that keeps the calculation perturbative. 
\end{itemize}

\section{Generalizing to the $N$-oscillator case}

	Now we generalize previous analysis for $N$ coupled oscillators. The Hamiltonian takes the following form, 
          \be \label{eq4.1}
          \mathcal{H}=\sum_{a=0}^{N-1}\frac{1}{2}\Big[p_{a}^2+\omega^2 x_{a}^2+\Omega^2(x_{a}-x_{a+1})^2+2\lambda\, x_{a}^4\Big].
          \ee
          We then do a discrete Fourier transformation to go to the normal mode coordinates. 
          \begin{align}
          \begin{split}
          &x_{a}=\frac{1}{\sqrt{N}}\sum_{k=0}^{N-1}\exp\Big(\frac{2\pi\,i\, k}{N}\,a \Big)\tilde x_{k},\\&
          p_{a}=\frac{1}{\sqrt{N}}\sum_{k=0}^{N-1}\exp\Big(-\frac{2\pi\,i\, k}{N}\,a \Big)\tilde p_{k}.
          \end{split}
          \end{align}
The Hamiltonian in terms of these variable becomes,
          \begin{align}
          \begin{split}\label{eq4.3}
          \mathcal{H}=&\sum_{a=0}^{N-1}\frac{1}{2}\Big[p_{a}^2+\omega^2 x_{a}^2+\Omega^2(x_{a}-x_{a+1})^2+2\lambda\, x_{a}^4\Big],\\&
          =\frac{1}{2}\sum_{k=0}^{N-1}\Big[|\tilde p_{k}|^2+\Big(\omega^2+4\Omega^2\sin^2\Big(\frac{\pi k}{N}\Big)\Big)|\tilde x_k|^2\Big]+\frac{\lambda}{N}\sum_{\alpha= N-k'-k_1-k_2\hspace{-0.25cm}\mod N, k',k_1,k_2=0}^{N-1}\tilde x_{\alpha}\tilde x_{k'}\tilde x_{k_1}\tilde x_{k_2}.          \end{split}
          \end{align}
    Here we have used the following facts,
  \be \label{omega}
  \tilde x_{k} =\tilde x_{k+N}, \quad \tilde x_{-k}=\tilde x^{\dagger}_{k}, \qquad \sum_{i=0}^{N-1}\exp\Big(-\frac{2\pi\,i(k-k')}{N}\Big)=N\delta_{k,k'}.
 \ee
   Also keeping in mind $\omega=m$ we define,
   \be
   \tilde  \omega_{i}^2=m^2+4\Omega^2\sin^2\Big(\frac{\pi\, i}{N}\Big),\quad  i=0,\cdots N-1.
   \ee      
 The ground state wavefunction  is (which is generalization of (\ref{eq3.17})),
 \begin{align}
 \begin{split}\label{eq4.5}
 \psi_{0,0,\cdots 0}(\tilde{x}_{0},\cdots\tilde x_{N-1})= \bigg(\frac{\tilde \omega_0\tilde \omega_1....\tilde \omega_{N-1}}{\pi^N}\bigg)^{\frac{1}{4}}\exp\Big(-\frac{1}{2}\sum_{k=0}^{N-1}\tilde \omega_k \tilde x_k^2+\lambda \psi^{1}\Big),
 \end{split}
 \end{align} 
 
        where,
         \begin{eqnarray*}
    \psi^1&=& \sum_{a=0 ;4a\,mod\,N=0 }^{N-1}B_{11}(a)+\frac{1}{2}\sum_{b,c=0 ;(2b+2c)\,mod\,N=0;b \neq c }^{N-1}B_{12}(b,c)+\sum_{d,e=0 ;(3e+d)\,mod\,N=0;e \neq d }^{N-1}B_{13}(d,e)\\
    & &+\frac{1}{2}\sum_{f,g,h=0 ;(f+2g+h)\,mod\,N=0 ; f \neq g \neq h}^{N-1}B_{14}(f,g,h)+\frac{1}{24}\sum_{i,j,k,l=0 ;(i+j+k+l)\,mod\,N=0 ; i \neq j \neq k \neq l}^{N-1}B_{15}(i,j,k,l),
\end{eqnarray*}
with
\begin{equation}\nonumber
     B_{11}(a)=-\frac{ \tilde x_a^4}{4 N \tilde \omega _a}-\frac{3  \tilde x_a^2}{4 N \tilde \omega
   _a^2}+\frac{9 }{16 N \tilde \omega _a^3},
\end{equation}

\begin{eqnarray*}
    B_{12}(b,c)&=&-\frac{3 \tilde  x_b^2 \tilde x_c^2}{N \left(\tilde \omega _b+\tilde \omega _c\right)}-\frac{3 
  \tilde  x_b^2}{2 N \tilde \omega _b \left(\tilde \omega _b+\tilde \omega _c\right)}-\frac{3 
  \tilde  x_c^2}{2 N \tilde \omega _c \left(\tilde \omega _b+\tilde \omega _c\right)}\\
   & &+\frac{3}{4 N \tilde \omega _b \tilde \omega _c \left(\tilde \omega _b+\tilde \omega
   _c\right)}+\frac{3}{4 N \tilde \omega _b^2 \left(\tilde \omega _b+\tilde \omega
   _c\right)}+\frac{3}{4 N \tilde \omega _c^2 \left(\tilde \omega _b+\tilde \omega
   _c\right)},\nonumber
\end{eqnarray*}

\begin{equation}\nonumber
     B_{13}(d,e)=-\frac{4 \tilde x_d \tilde x_e^3}{N (\tilde \omega _d+3  \tilde \omega _e)}-\frac{12 \tilde x_d
   \tilde x_e}{N \left(\tilde \omega _d+\tilde \omega _e\right) \left(\tilde \omega _d+3\tilde  \omega
   _e\right)} ,
\end{equation}

\begin{equation}\nonumber
     B_{14}(f,g,h)=-\frac{12 \tilde  x_f \tilde  x_g^2 \tilde x_h}{N \left(\tilde \omega _f+2 \tilde \omega _g+\tilde \omega
   _h\right)}-\frac{12\tilde  x_f \tilde x_h}{N \left(\tilde \omega _f+\tilde \omega _h\right)
   \left(\tilde \omega _f+2\tilde  \omega _g+\tilde \omega _h\right)},
\end{equation}

\begin{equation}\nonumber
     B_{15}(i,j,k,l)= -\frac{24 \tilde x_i\tilde  x_j \tilde x_k\tilde  x_l}{N \left(\tilde \omega _i+\tilde \omega _j+\tilde \omega
   _k+\tilde \omega _l\right)}.
\end{equation}
 Again we should keep in mind that (\ref{eq4.5}) is only valid upto $\mathcal{O}(\lambda).$ We will expand the complexity upto $\mathcal{O}(\lambda).$ (\ref{eq4.5}) can be recast into the following form,
          
     \be\label{eq4.6}
           \psi^{s=1}_{0,0,\cdots 0}(\tilde{x}_{0},\cdots\tilde x_{N-1})\approx\exp\Big(-\frac{1}{2}v_{a}.A^{s=1}_{ab}.v_{b}'\Big).
           \ee  \label{eq4.7}

As seen from equation (\ref{eq4.5}), the ground state has the sum of following type of terms within the exponential : $\tilde x_i^2, \tilde x_i \tilde x_j, \tilde x_i^4, \tilde x_a^2 \tilde x_b^2, \tilde x_f \tilde x_g^2 \tilde x_h, \tilde x_d \tilde x_e^3, \tilde x_i \tilde x_j \tilde x_k \tilde x_l $. Given this, the choice of basis is not unique.  But there is a minimal choice such that, the basis for $A(s=1)$ is

\be
\vec v=\{\tilde x_0, \tilde x_1,....,\tilde x_{N-1},\tilde x_0^2,\tilde x_1^2,....,\tilde x_{N-1}^2,..,\tilde x_i \tilde x_j,...\}.
\ee
The total number of terms in this basis are $ N+ \frac{N(N-1)}{2}= \frac{N(N+1)}{2}$ and for large $N$ it grows as $N^2.$ For more discussions on the counting of the basis refer to the Appendix~(B). Given this  choice of basis the matrix $A$ takes a block diagonal form
\be
           A^{s=1}=\left(
\begin{array}{cc}
 A_1 & 0 \\
 0 & A_2 \\
\end{array}
\right).
          \ee
\\
 Now again there are several comments are in order. 
 \begin{itemize}
 \item{
 The elements of the block $A_1$ constitutes $-2 \times$ coefficient of terms of type $\tilde x_a^2, \tilde x_i \tilde x_j$ within the exponential in the target state. The form of the matrix $A_1$ is unique once we fix the target state (\ref{eq4.5}). We will at times refer to the block $A_1$ as the \textit{`unambiguous' block.}} The reason for this will be clear from the discussion below.
 \item {The elements of the block $A_2$ constitutes $-2 \times$ coefficient of terms of type  $$\tilde x_i^4, \tilde x_a^2 \tilde x_b^2, \tilde x_f \tilde x_g^2 \tilde x_h, \tilde x_d \tilde x_e^3, \tilde x_i \tilde x_j \tilde x_k \tilde x_l,$$ within the exponential in the target state. This block ($A_2$) is not uniquely fixed even after the target state is fixed to be (\ref{eq4.5}). This is because elements corresponding to $\tilde x_a^2 \tilde x_b^2$ in the target state, can be either put in the diagonal entry of the matrix $A$ corresponding to the basis element $\tilde x_a \tilde x_b$ or as an off diagonal element corresponding to the basis entries $\tilde x_a^2$ and $\tilde x_b^2$. This sort of ambiguity also arises in the entries corresponding to $\tilde x_i \tilde x_j \tilde x_k \tilde x_l$. These are always off diagonal, but can be put in the target matrix in more than one way. In the most general target matrix such elements are distributed among all possible entries of the $A_2$ matrix, such that the sum of all entries add up to the -2$\times$ coefficient under consideration. Hence we will sometimes refer to the block $A_2$ as the \textit{`ambiguous block'}. }

\item{ Given the freedom in choice of $A_2$ we now make another choice for this rearrangement inside $A_2$ such that the determinant of $A_2$ is always nonzero i.e all the eigenvalues of this matrix is nonzero. This is absolutely necessary for our geodesic analysis for which  the determinant of target and reference matrix has to be non zero.}
 \end{itemize}
 
Next we have to  choose the reference state. Again as described in the previous section, a desirable property of the reference state is that it should not contain any entanglement in the original coordinates. So we generalize the reference state mentioned in (\ref{refstate}) for arbitrary $N$ in the following way. 
\begin{align}
\begin{split} \label{refstate1}
\psi^{s=0}(x_1,x_2,\cdots,x_{n})=\mathcal{N}^{s=0}\exp\Big[-\frac{\tilde \omega_{ref}}{2}\Big(\sum_{i=0}^{N-1}(x_i^2+\lambda_0\,x_i^4)\Big)\Big].
\end{split}
\end{align}
  This again can be recast in the following way (after going to the normal mode coordinate), 
\be
\psi^{s=0}(\tilde x_1, \tilde x_2,\cdots, \tilde x_{n})=\mathcal{N}^{s=0}\exp\Big[-\frac{\tilde \omega_{ref}}{2}\Big(v_{a}.A^{s=0}_{ab}.v_{b}\Big)\Big]
\ee
where, 
 \be\label{eq4.9}
 A^{s=0}=\left(
\begin{array}{cc}
 \tilde \omega_{ref}I_{N\times N} & 0 \\
 0 &\tilde \omega_{ref}\lambda_{0}\,I_{k\times k} \\
\end{array}
\right). 
 \ee
 
 Dimension of the identity matrix $I_{k \times k}$ is same as that of the dimension of $A^{(2)}.$ It is evident that $A^{s=0}$ also decomposes in terms of an unambiguous block and an ambiguous block. Again there will be all those ambiguities regarding the rearrangements of the elements inside this ambiguous  block of  the reference state as discussed previously in the context of the reference state. We also make a choice such that the determinant of $A^{s=0}$ becomes non zero as in the $N=2$ analysis. \\
 Armed with this reference state let us now discuss the complexity functional. We will use the following type of functional:
\be \mathcal{F}_{\kappa}= \sum_{I} p_{I} \Big|Y^{I}(s)\Big|^{\kappa}.
\ee 
Then the complexity will be given,
\be \mathcal{C}_{\kappa}= \int_{0}^{1}\, ds\, \mathcal{F}_{\kappa}.
\ee \label{eq4.10}

Note that this is more general than the functional ($\mathcal{F}_{2}$) that we used in the previous section. As in the two oscillator case, the target ($ A^{s=1}$) and reference matrix ($ A^{s=0}$)   commute with each other and they can be  simultaneously diagonalized. So again the complexity will be given by the ratio of the eigenvalues of the reference and target state as stated below \footnote{ For arbitrary $N, $ $\mathcal{F}_{2}$ will be given by, $\mathcal{F}_{2}= \frac{1}{2}\Big(\sqrt{\sum_{i=1}^{N-1} \log^2\Big(\frac{\lambda_i^{(1)}}{\tilde\omega_{ref}}\Big)+\sum_{j} \log^2\Big(\frac{\lambda_j^{(2)}}{\tilde\omega_{ref}}\Big)}\Big).$}, 
\begin{align}
\begin{split} \label{func}
\mathcal{C}_{\kappa}=\frac{1}{2^{\kappa}}\sum_{i=0}^{N-1}\Big|\log\Big(\frac{\lambda_{i}^{(1)}}{\tilde \omega_{ref}}\Big)\Big|^{\kappa}+\mathcal{A} \sum_{j}\Big|\log\Big(\frac{\lambda_{j}^{(2)}}{h_{j}\,\tilde \omega_{ref}\lambda_{0}}\Big)\Big|^{\kappa},
\end{split}
\end{align}
 $\lambda_i^{(1)}$ are the eigenvalues coming from the \textit{`unambiguous' block} and $\lambda_{j}^{(2)}$  are the eigenvalues coming from the \textit{`ambiguous' block}. $h_{j}$ are some numbers. The number of $\lambda_{j}^{(2)}$ is the same as the dimension of the unambiguous block and for large $N$ it grows as $N^2.$
 For $\kappa=1$ the complexity functional  (\ref{func}) can be rewritten in the following way.
  \begin{align}
 \begin{split}
\mathcal{C}_{\kappa=1}=\frac{1}{2}\Big(\log \Big(\frac{\det (A^{(1) s=1 })}{\det (A^{(1) s=0} )}\Big )+\mathcal{A}\log\Big(\frac{\det (A^{(2) s=1})}{\det (A^{(2) s=0})}\Big)\Big).
 \end{split}
 \end{align}
  All the individual eigenvalues coming from the unambiguous  blocks are positive both for the target and reference states. So the first ratio is automatically positive.  But not all eigenvalues  coming from the ambiguous blocks are positive. Some of them turns out to be negative. As we know there are several ambiguities inside this block. However, we cannot find any choice for the arbitrary parameters such that all the eigenvalues coming from this block will always be positive for all values of $N.$ Similar problem persists for the ambiguous  part of the reference  block. But we can always check  that for a given $N,$ there are always some choices for the rearrangement inside this ambiguous  block such that  this ratio  $\frac{\det (A^{(2) s=1})}{\det (A^{(2) s=0})}$ is always positive.  Hence the $\mathcal{C}_{\kappa=1}$ is well defined. In fact one of the Schatten norms $\mathcal{F}_{p}$ with $p=2$ is also well defined for our case. For our case that is just $\sqrt{\mathcal{C}_{\kappa=1}}.$ It seems that  at this moment only these two measures are the only two measures  that are well defined for our case. We will use $\mathcal{C}_{\kappa=1}$ only for all the subsequent discussions for simplicity. It will be an interesting problem to investigate  the other measure but we will leave it for future investigations.\\
 Next we evaluate this complexity functional. As evident from (\ref{func})  this complexity has two pieces.
 \be
 \mathcal{C}_{\kappa}=\mathcal{C}^{(1)}_{\kappa}+\mathcal{C}^{(2)}_{\kappa},
 \ee
 where, $\mathcal{C}_{\kappa}^{(1)}$ comes from the unambiguous piece and $\mathcal{C}_{\kappa}^{(2)}$ comes from ambiguous piece including the extra penalty factor $\mathcal{A}.$ In the next section we evaluate these two pieces separately.

\subsection{Evaluation of complexity functional and continuous limit}
Before we proceed to compute the complexity functional we like to reinstate the factor of $M$ that we have set to one from (\ref{eq3.5}) onwards. With this factor reinstated, the Hamiltonian takes the following form,
\begin{align}
\begin{split} \label{ref1}
\mathcal{H}=\frac{1}{M}\sum_{\vec n}\Big\{\frac{P(\vec n)^2}{2}+\frac{1}{2}M^2\Big[\omega^2 X(\vec n)^2+\Omega^2 \sum_{i} (X(\vec n)-X(\vec n-\hat x_{i}))^2+2 \lambda\, X(\vec n)^4\Big]\Big\}.
\end{split}
\end{align}
Now the overall factor of $\frac{1}{M}$ doesn't change the form of the ground state wavefunction. Only now there is a nontrivial factor of $M^2$ infront of the $x^2$ and $x^4$ part of the Hamiltonian. It will just scales various quantities,
$$\tilde \omega_{i_k}\rightarrow \frac{\tilde \omega_{i_k}}{\delta},\quad \Omega\rightarrow \frac{\Omega}{\delta},\quad \lambda\rightarrow\frac{\lambda}{\delta^2}, \quad \tilde \omega_{ref}\rightarrow \frac{\tilde \omega_{ref}}{\delta},\lambda_{0}\rightarrow \frac{\lambda_0}{\delta}.$$

In light of this we have the general formula for $\lambda^{(1)}_{i_k}$ as shown below,
\begin{align}
\begin{split}\label{eq4.12}
\lambda_{i_k}^{(1)}=&\,\,\frac{\tilde \omega_{i_k}}{\delta}+\frac{3\,\lambda}{2 N}\Big(\frac{2}{\tilde \omega_{i_k}(\tilde \omega_{i_k}+\tilde \omega_{N-i_k})}+\frac{2}{\tilde\omega_{i_k}(\tilde\omega_{i_k}+\tilde \omega_{\frac{N}{2}-i_k})}\Big),\quad  N:even\\&\frac{\tilde \omega_{i_k}}{\delta}+\frac{3\,\lambda}{2 N}\Big(\frac{2}{\tilde \omega_{i_k}(\tilde \omega_{i_k}+\tilde \omega_{N-i_k})}\Big),\quad  N: odd
\end{split}
\end{align}
$i_k$ goes from $0$ to $N-1$ for all $k$ from $1$ to $d-1.$  $N$ denotes the number of lattice points in each of the spatial directions.  Then the $d-1$ dimensional spatial volume is given by $ L^{d-1}=( N\,\delta)^{d-1}.$
Using (\ref{eq4.12}) we get the contribution to the complexity coming from the  unambiguous block.  
\begin{align}
\begin{split}\label{eq4.13}
\mathcal{C}^{(1)}_{\kappa}=&\frac{1}{2^{\kappa}}\sum_{k=1}^{d-1}\Big[\sum_{i_k=0}^{N-1}\Big|\log\frac{\tilde \omega_{i_k}}{\tilde \omega_{ref}}\Big|^\kappa+\frac{3\lambda\,\kappa\, \delta}{2\,N}\,\sum_{i_k=0}^{N-1}\,\frac{1}{ \tilde \omega_{i_k}^3}\Big|\log(\frac{\tilde \omega_{i_k}}{\tilde \omega_{ref}})\Big|^{\kappa-1}\Big]+\mathcal{O}(\lambda^2).
\end{split}
\end{align}
From now on we will focus on the $\kappa=1$ case. Also, we will take the continuous limit i.e $N\rightarrow \infty$ and $\delta\rightarrow 0$ such that $N\delta $ is finite. We are mostly interested in extracting the leading divergent and finite  terms of $\mathcal{C}^{(1)}_{\kappa=1}.$  From now onwards we will write every expression in terms of $\hat \lambda= 24\,\delta^{d}\, \lambda $ and $V= (N \delta)^{d-1},$ instead of $\lambda,\, N$ and $\Omega=\frac{1}{\delta}.$ Also we quote here an useful relation which is the generalization of (\ref{omega}) for arbitrary $d.$
\be
\sum_{k=1}^{d-1}\tilde \omega_{i_{k}}^2=m^2+\frac{4}{\delta^2}\, \sum_{k=1}^{d-1}\sin^2\Big(\frac{\pi\,i_{k}}{N}\Big),
\ee
with each of the $i_{k}$ goes from $0$ to $N-1$ for all $k.$
\subsection*{d=2}
Using the general method described in Appendix~(C), we arrive at the total expression for the complexity given below: 
\begin{align}
    \begin{split}
        \mathcal{C}^{(1)}_{\kappa=1}=&\frac{V}{2\,\delta}\log\left(\frac{m}{\tilde\omega_{ref}}\right)+ \frac{V}{2\delta} \log \left(\frac{1}{2} \sqrt{\frac{4}{(m\,\delta)^2}+1}+\frac{1}{2}\right)+ \frac{\hat\lambda}{8\,\pi\,m^2}\frac{E\left(\frac{4 }{(m\,\delta)^2+4 }\right)}{\sqrt{(m\,\delta)^2+4 }}. \\
    \end{split}
\end{align}
Here $E(k)$ is the elliptic function of the second kind. This captures the exact $\delta, m$ dependence upto leading order in $\hat \lambda$. 
We expand in terms of $\delta$ and that gives, 


\begin{equation} \label{d=2}
    \mathcal{C}^{(1)}_{\kappa=1}=\frac{V}{2 \delta}\log\left(\frac{1}{\tilde\omega_{ref}\delta}\right)+\frac{V}{\delta}\Big(a_{1}+c_1\,(m\,\delta)+c_3(m\,\delta)^3+\mathcal{O}((m\,\delta)^5)\Big)+ \frac{\hat \lambda\,\delta^2}{ 16} \Big(\frac {f_1}{(m\delta)^2}+ f_{1,log}\,\log (m\, \delta)+f_0+\cdots\Big),
    \end{equation}\\
where $a_1=0$, $c_1=1/4, c_3=-\frac{1}{96}, f_{1}=\frac{1}{\pi},f_{1,log}=-\frac{1}{8\,\pi},f_0=0.02.$ 

We note that, in comparison with the numerically computed free theory part- expression (E.12) in \cite{jm}, the exact free theory result has all terms except the constant $a_0$ and the $c_1 V\, m \log\left(\frac{m}{\tilde\omega_{ref}}\right)$ terms.
 
 \subsection*{d=3}
 
 Using the results in Appendix~(C), we find that for $d=3:$

\begin{align}
    \begin{split}\label{d=3}
        \mathcal{C}^{(1)}_{\kappa=1}=&\frac{V}{2\,\delta^2}\log\left(\frac{1}{ \tilde\omega_{ref}\, \delta}\right)+ \frac{V}{\delta^2}\Big(a_2+b_2\, (m\,\delta)^2\, \log(m\delta)+c_2 (m\,\delta)^2+\mathcal{O}( (m\delta)^3)\Big)
      +f_1\,\frac{\hat{\lambda }}{16} \frac{V^{1/2}}{(m\,\delta)}+\cdots,
    \end{split}
\end{align}
where, $ a_2= 0.29,b_2=0.075, c_2=-0.04$  and $f_1=0.16$. 
Here we have retained only the leading $\frac{1}{(m\delta)}$ term which arises from the integral (\ref{A6}) for the interacting sum.
Note that there is a $\log (m\delta)$ factor multiplying $V\, m^2$ contrasted to the fact that there is no such logarithmic term multiplying $(V \, m)$ term in (\ref{d=2}). Furthermore, unlike \cite{jm}, the free theory result that our analysis gives is manifestly proportional to volume and only in the interacting part is there a breakdown in $V$ scaling, since it is proportional to $V^{1/2}$. Notice that compared to $d=2$, the interaction part has an extra $1/\delta$ dependence suggesting that as $d$ increases, for fixed $\hat\lambda$ complexity will increase.

\subsection*{General d}
  For arbitrary $d$, as we argue in Appendix~(C), we have
\begin{align}
    \begin{split} \label{d=4}
        \mathcal{C}^{(1)}_{\kappa=1}=&\frac{V}{2\,\delta^{d-1}}\log\left(\frac{1}{ \tilde\omega_{ref}\, \delta}\right)+  \frac{V}{\delta^{d-1}}\left(a_{d-1}+\log(m\delta) \left[\sum_{k=2}b_k (m\delta)^k\right]+\sum_{k=1} c_k (m\delta)^k\right) \\&+\frac{\hat\lambda}{16} \delta^{6-2d} V^{\frac{d-2}{d-1}}\left(f_1 \{(m\delta)^{d-4}|_{d\neq4}+\log(m\delta)|_{d=4}\}+f_0+\cdots\right).
    \end{split}
\end{align}
Using the results in Appendix~(C) we find $a_3= 0.41, a_4=0.49, a_5=0.55$. In principle we can also fix the $b_i, c_i$'s for general dimensions, but we will not attempt to do it here. Also we note from (\ref{d=2}) that the $d=2$ case is  somewhat special. For $d\geq 3$ there is no $c_1$ term, it is only nonzero for $d=2.$  Further we can determine $c_2=-0.02, b_2=0$ for $d=4$. We tabulate $f_0, f_1$ below after dividing  them by a factor of $\Gamma(\frac{3}{2})$ \footnote{The values for $d=3.99$ are perfectly consistent with $d=4$ as can be checked by writing $1/(m\delta)^{0.01}\sim 1-0.01\log (m\delta) $.}:
\begin{table}[h]
\begin{center}
\begin{tabular}{|l|c|c|c|c|}
\hline
$\Gamma(\frac{3}{2}) f_i$ & d=3 & d=3.99 & d=4 & d=5 \\
$\Gamma(\frac{3}{2}) f_0$ & -0.001 & -4.75 & 0.07 & 0.07\\
$\Gamma(\frac{3}{2}) f_1$ & 0.14 & 4.83 & -0.05 & -0.03 \\
\hline
\end{tabular}
\end{center}
\end{table}
 
 Note the flip in signs for $d<4$ as opposed to $d>4$. 
 
 \subsection*{d=$4-\epsilon$}
The general method outlined in Appendix~(C) enables us to extract information for any dimension, not just integer dimensions. For instance, using the integral form for the free theory sum, we can extract $a_{d-1}$ as a function of $d$. This is shown in Figure~(1).
\begin{figure}[h]
\begin{center}
\vskip 2pt
\resizebox{250pt}{140pt}{\includegraphics{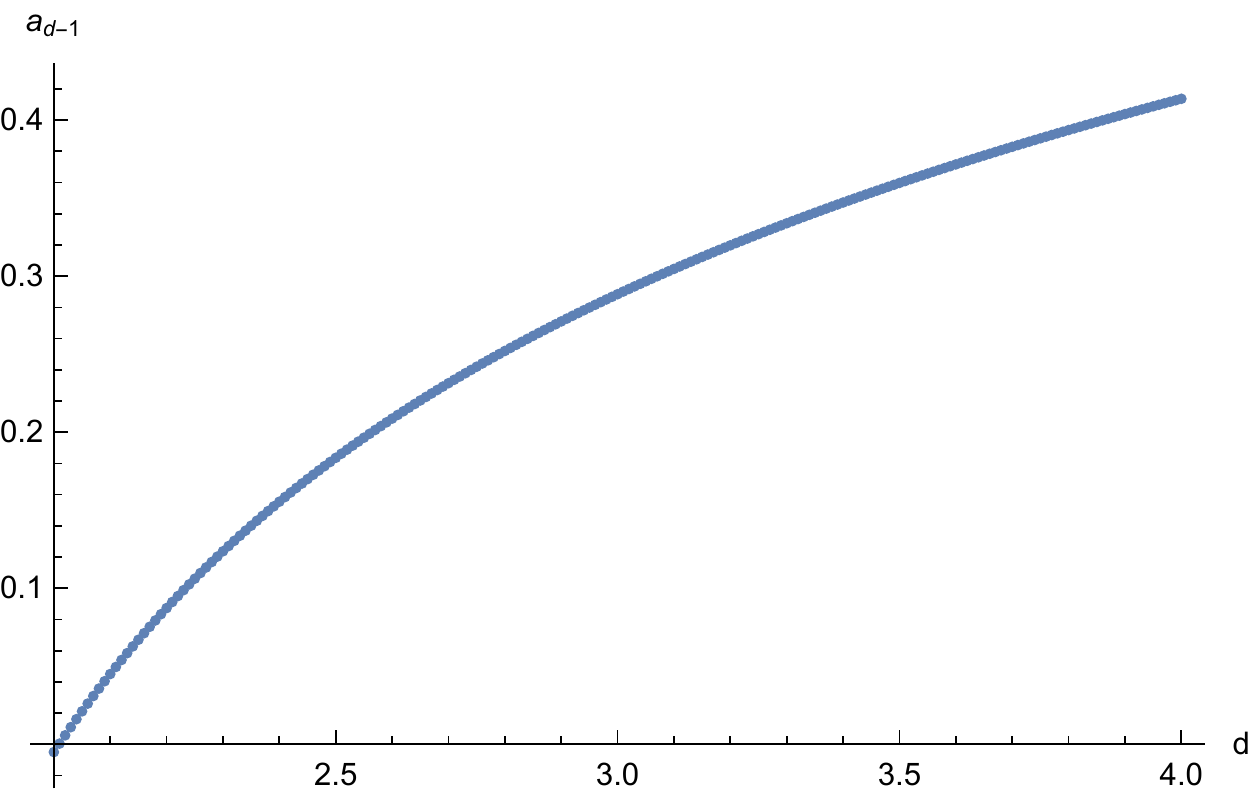}}
\end{center}
\caption{ $a_{d+1}$ as a function of $d$.}
\end{figure}
We can also evaluate the interacting sum as a function of $\hat m\equiv m \delta$ for various dimensions. This is illustrated in Figure~(2). The plot is consistent with our findings above--for instance, it shows a divergence for $\hat m=0$ for $d<4$ while it approaches a fixed value for $d>4$. 
\begin{figure}[h]
\begin{center}
\vskip 2pt
\resizebox{250pt}{140pt}{\includegraphics{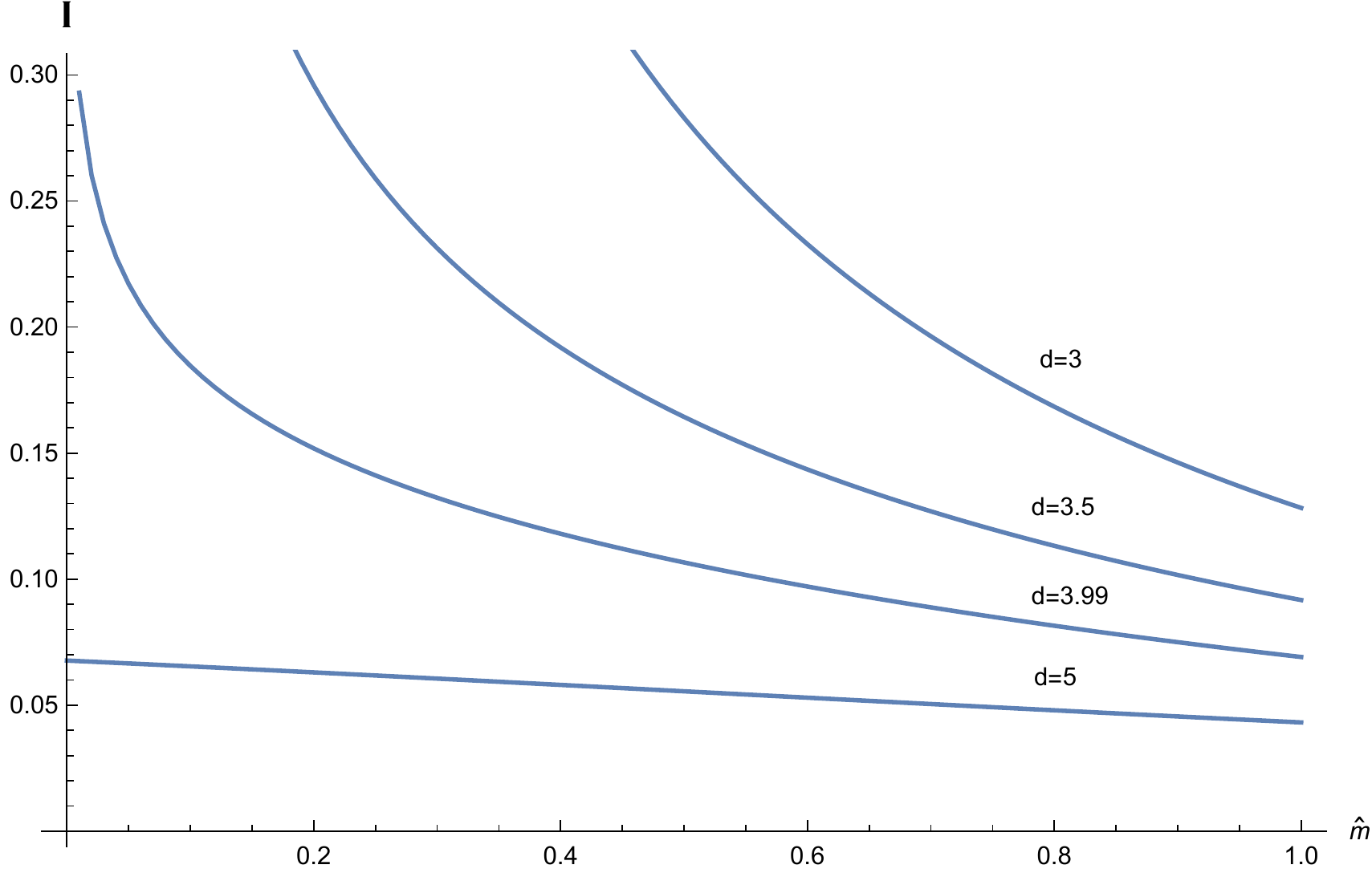}}
\end{center}
\caption{The interacting sum is $\frac{\hat\lambda}{8\,\sqrt{\pi}} \delta^{6-2d} V^{\frac{d-2}{d-1}}$ times the y-axis.}
\end{figure}

Thus we can systematically study the epsilon expansion. At leading order since $\hat \lambda_{*}=\frac{16\pi^2}{3}\epsilon$ at the fixed point, the result is somewhat trivial since we can replace the $f_0,f_1$ by the $d=4$ values. However this procedure will prove to be useful when one computes the next order in perturbation.
 \subsection{Complexity in terms of renormalized parameters}
 Up to now we have given expressions involving the bare parameters. In order to extract physics, we will need to rewrite the expressions above in terms of the renormalized quantities following\cite{smit, phi4book}. For the mass we have \cite{smit},
 \be
 (m\,\delta)^2= (m_{R}\,\delta)^2-\frac{ \lambda_{R}\,\delta^{4-d}}{2}I (m_R\delta)+O(\lambda_R^2)\,.
 \ee
 $m_R$ is the renormalized mass and $\lambda_R$ is the renormalized coupling defined at zero momentum. A running renormalized coupling can also be defined at finite momentum $\mu$ and since we are interested only in leading order in the coupling, this amounts to simply replacing $\lambda_R$ by $\lambda_R(\mu)$. Here,
 \be
 I(m_{R}\delta)=\prod_{i=1}^{d}\Big[\int_{-\pi}^{\pi}\frac{dl_i}{(2\pi)}\Big]\,\frac{1}{(m_R\,\delta)^2+4\,\sum_{i=1}^{d}\sin^2(\frac{l_i}{2})}\,.
 \ee
 For $d=2$ we get,
 \be
 (m \,\delta)^2=(m_{R}\,\delta)^2-\frac{ \lambda_{R}\,\delta^2}{2}\left[C_0-2\,C_1 \log ( m_{R}\,\delta)- C_2 (m_{R}\,\delta)^2+\frac{1}{32\,\pi}(m_R\,\delta)^2\log ( (m_R\,\delta)^2)+\mathcal{O}((m_{R}\,\delta)^4)\right].
 \ee
Here, $C_0= 0.28, C_1=0.08,C_2=0.02.$ For other dimensions we have used the ``{\it hopping expansion}" method of \cite{phi4book}.\par 
 For $d\geq 3$ we get,
 \be
(m \,\delta)^2=(m_{R}\,\delta)^2-\frac{\lambda_R\,\delta^{4-d}}{2}\left[C_0- C_2 (m_{R}\,\delta)^2+\frac{1}{16 \pi^2} (m_{R}\,\delta)^2\log ((m_{R}\,\delta)^2)|_{d=4}+\mathcal{O}((m_{R}\,\delta)^4)\right].
\ee 
Note that for $d=4$ there is an extra log term. Also we tabulate values of $C_0$ and $C_2$ for various dimensions. 
\begin{table}[h]
\begin{center}
\begin{tabular}{|l|c|c|c|c|}
\hline
$C_i$ & d=3 & d=3.99&d=4 & d=5 \\
$C_0$ & 0.21 &0.15 & 0.15&0.11  \\
$C_2$ &0.06 &0.03&0.03 &0.015\\
\hline
\end{tabular}
\end{center}
\end{table}

 From this  we note that $C_0$ and $C_2$ are always positive for all $d$. At this order we simply have $\hat \lambda_0=\lambda_{R},$ where $\lambda_R$ is the renormalized coupling. Now given these two expressions, we get the following.
\begin{align}
\begin{split} 
 d=2:\,\,
\mathcal{C}_{\kappa=1}^{(1)}=&\frac{V}{2 \delta}\log\left(\frac{1}{\tilde\omega_{ref}\delta}\right)+\frac{V}{\delta}\Big(a_{1}+c_1\,(m_{R}\,\delta)+\cdots\Big) -{\lambda_{R}\,\delta^2}I(m_R\delta) \frac{V}{2\delta}\Big(\frac{c_1}{2m_R\delta}+\cdots\Big)\\&+\frac{\lambda_R\delta^2}{16} \left(\frac{f_1}{m_R^2\delta^2}+f_{1,log}\log(m_R\delta)+f_0\right)+\cdots,\\&\approx \frac{V}{2\delta}\left[\log\left(\frac{1}{\tilde\omega_{ref}\delta}\right)+2a_1-\lambda_R \delta^2\frac{C_0-2C_1\log(m_R\delta)}{2 m_R\delta}c_1\right]+\cdots\,.
\end{split}
\end{align}
\begin{align}\
\begin{split}
d\geq 3: \,\,\mathcal{C}_{\kappa=1}^{(1)}=& \frac{V}{2\,\delta^{d-1}}\log\left(\frac{1}{ \tilde\omega_{ref}\, \delta}\right)+  \frac{V}{\delta^{d-1}}\left(a_{d-1}+\log(m_R\,\delta) \left[\sum_{k=2}b_k (m_R\,\delta)^k\right]+\sum_{k=2} c_k (m_R\,\delta)^k\right) \\&-\frac{V}{4\,\delta^{d-1}}(\lambda_{R}\,\delta^{4-d})I(m_R\delta)\Big[\sum_{k=2}\,(m_R \delta)^{k-2} (k b_k\log(m_{R}\,\delta)+b_k +k c_k)\Big]\Big)\\&+\frac{\lambda_R}{16} \delta^{6-2d} V^{\frac{d-2}{d-1}}\left(f_1 \{(m_R\,\delta)^{d-4}|_{d\neq4}+\log(m_R\,\delta)|_{d=4}\}+f_0\right)+\cdots\\ &\approx \frac{V}{2\delta^{d-1}}\left[\log\left(\frac{1}{\tilde\omega_{ref}\delta}\right)+2a_{d-1}-\lambda_R \delta^{4-d} C_0 (c_2+b_2\log(m_R\delta)+\frac{b_2}{2})\right]\\ & +\frac{\lambda_R}{16} \delta^{6-2d} V^{\frac{d-2}{d-1}}\left(f_1 \{(m_R\,\delta)^{d-4}|_{d\neq4}+\log(m_R\,\delta)|_{d=4}\}+f_0\right)+\cdots
\end{split}
\end{align} 
  In both the expressions above, we have indicated the dominant terms in the small $\delta$ limit, keeping $\lambda_R, m_R$ finite. For $d\geq 4$, $b_2=0$. Also note that the genuine extra contributions that came from the first block which were proportional to $f_0, f_1$ give a fractional dependence on the volume and are subleading in the large volume limit. Now it is obvious that since the $\lambda_R$ dependence at leading order in $V$ is proportional to $\delta^{4-d}$, perturbation theory will break down for $d>4$. This is expected from the RG picture where $d>4$ has a stable Gaussian fixed point but an unstable Wilson-Fisher fixed point. In order to isolate the effect of the interaction we can define (for $d>3$)
  \be
  \Delta \mathcal{C}_{\kappa=1}^{(1)}=\mathcal{C}_{\kappa=1}^{(1)}-\mathcal{C}_{\kappa=1}^{(1)}|_{\lambda_R=0}\approx  -\frac{V}{2\delta^{d-1}}\lambda_R\delta^{4-d} C_0 c_2\,.
  \ee
  In other words we are asking what is the change in complexity when we go from the free theory with mass parameter $m_R$ to the interacting theory with the same mass parameter.
At the Wilson-Fisher fixed point in the epsilon expansion $\lambda_{R*}=\frac{16\pi^2}{3}\epsilon$ and to leading order we use $c_2=-0.02, C_0\approx 0.15$, the 4-dimensional value. This means that interaction has slightly increased the complexity at the fixed point compared to what happens at the Gaussian fixed point. For both fixed points $m_R=0$. The sign of $c_2$ will turn out to be important when discussing the consequences from the flow equation.

\subsection{Comments about $\mathcal{C}^{(2)}_{\kappa}$ and structure of penalty factor}
As discussed around  equation (\ref{func}) there is a second contribution coming from the eigenvalues of ``\textit{ambiguous}" block ($\lambda^{(2)}_{j}$). As the name suggests and also from the earlier discussions there are many ambiguities. Nonetheless we will discuss the general structure of the contributions coming from this block to the complexity expression ($\mathcal{C}^{(2)}_{\kappa=1}$). We present all the expressions in terms of the renormalized quantity.
 
\subsection*{ Structure of the $\lambda^{(2)}_{j}$ eigenvalues}
$\lambda^{(2)}_{j}$ eigenvalues, unlike the $\lambda^{(1)}_{j}$'s,  do not seem to have a straightforward generalized expression for $N$-oscillator case, even after choosing a particular rearrangement for the second block $A_2.$ But all elements of the $A_2$ matrix are of the form $A_2[m,n]$ = $ \frac{a_{mn}\,\lambda_R\,\delta^{-d}}{V^{\frac{1}{d-1}}\, f(\tilde\omega_i)}$ for $i$ $\in$ $\{0,1,..N-1\}$ and $ f(\tilde\omega_i)$ are the linear functions of $\tilde \omega_{i}.$ Given this  the eigenvalues $\lambda^{(2)}_{j}$ takes the general form \be \label{o2} \lambda^{(2)}_j= \frac{b_j\,\lambda_R\,\delta^{-d}}{V^{\frac{1}{d-1}}\, g(\tilde\omega_i)  },\ee for $j$ $\in$ $\{0,1,..(\text{Dim }A_2)-1 \}$, $i$ $\in$ $\{0,1,..N-1\}$. Here $ g(\tilde\omega_i)$ has the dimension of $m.$ 


Since we are interested in the divergence of the leading term that contributes to complexity, we will assume a simpler form of the eigenvalues (based on above points), 
\begin{equation} \label{o1}
    \lambda^{(2)}_j= \frac{b_j\,\lambda_R\,\delta^{-d}}{V^{\frac{1}{d-1}}\,\tilde\omega_j };\hspace{2cm}j \in \{0,1,..,(\text{Dim }\mathcal{A}_2)-1 \}.
\end{equation}
 Now we have to keep  in mind that,  $$\tilde\omega_i=\tilde\omega_{N+i}=\tilde\omega_{N-i}=\tilde\omega_{-i}.$$ Now the Dimension of the block $A_{2}$ grows as $N^2$ for large $N.$ But there are only $N$ number of $\tilde \omega_{i}.$ So given the form of $\lambda^{(2)}_{j},$ either in the equation (\ref{o1}) or in (\ref{o2}), there will be only $N$ number of $\lambda_{j}^{(2)}$ eigenvalues each with degeneracy $N.$ This argument can be extended straightforwardly   for any dimensions $d.$   

\subsection*{ Cost function and the penalty factor}

We will use the cost function as mentioned in (\ref{func}). We will concentrate on $\mathcal{C}_{\kappa=1}^{(2)}.$  We will now make the following choice for the penalty factor\footnote{Calling this a penalty factor is a bit of a misnomer since we are in fact unpenalising the gates to have a perturbative behaviour for complexity.} $\mathcal{A}$ as mentioned in (\ref{func}):  
\be   \label{it}
\mathcal{A}= (\lambda_R\, \delta^{4-d})^{\mu} \delta^{-\nu}\, V^{\frac{\nu}{d-1}}.
\ee
$\mathcal{A}$ should be dimensionless and for the moment $\mu$ and $\nu$ are arbitrary but integers. But we will soon make choice based on some physical arguments. With this choice we get, 

\begin{equation}\label{c1}
    C_{\kappa=1}^{(2)}=\frac{(\lambda_R\, \delta^{4-d})^{\mu} \delta^{-\nu}\, V^{\frac{\nu}{d-1}}}{2} \sum_{j}\Big|\log\Big(\frac{\lambda_{j}^{(2)}\delta^2}{\tilde \omega_{ref}\lambda_{0}}\Big)\Big|.
\end{equation}\\
In the original basis the reference matrix is chosen to be the one with eigenvalues $\lambda_0\, \tilde\omega_{ref}$ coming from the ``{\it ambiguous"} block of the target state. This ensures that in the diagonal basis, the eigenvalues are of the form  
 $h_{i_k}  \lambda_0\,\tilde\omega_{ref},$ where $h_{i_k}$ are some  constant numbers. We absorb this  $h_{i_k}$ into the $b_{i_k}$.  Also, to make the $\hat \lambda\rightarrow 0$ limit well defined, we assume as discussed previously, $\lambda_{0} $ to be proportional to $ \lambda_R$ so that the $ \lambda_R$ dependence inside the logarithm will cancel out. 
Given this, and the form of $\mathcal{A},$  we can easily see that $ \lambda_R \rightarrow 0$ limit is well defined. Now this leaves us with two possibilities for choosing $\lambda_0.$
\be \label{zerol}
\lambda_0=a\,\lambda_R\, \tilde\omega_{ref}^{d-3}, \quad \text{or}\, \quad \lambda_0=a\,\lambda_R\, m_{R}^{d-3}
\ee
Putting all these pieces together (for general dimensions) and reinstating all the necessary factors of $\delta$ we get, 
\begin{equation}\label{intm}
    C_{\kappa=1}^{(2)}=\frac{(\lambda_R\, \delta^{4-d})^{\mu} \delta^{-\nu}\, V^{\frac{\nu}{d-1}}}{2}
    \sum_{k=1}^{d-1}\sum_{i_k=0}^{(\text{Dim }A_2)-1}\Bigg|\log\left(\frac{b_{i_k}\, \lambda_{R}\,  \delta^{2-d}}{V^{\frac{1}{d-1}}\,\lambda_0\, \tilde\omega_{i_{k}} \tilde\omega_{ref}}\right)\Bigg|.
\end{equation}
Now we use the fact these $\lambda^{(2)}_{j}$ eigenvalues are degenerate. For each $i_{k}$ where $k=1,\cdots d-1,$ there are $N$ of these eigenvalues with degeneracy $N.$ Using this fact we get,
\begin{equation}\label{intm1}
    C_{\kappa=1}^{(2)}=\frac{V}{\delta^{d-1}}\frac{(\lambda_{R}\, \delta^{4-d})^{\mu} \delta^{-\nu}\, V^{\frac{\nu}{d-1}}}{2}\sum_{k=1}^{d-1}\sum_{i_k=0}^{N-1}\log\left(\frac{b_{i_k}\, \lambda_R\,  \delta^{2-d}}{V^{\frac{1}{d-1}}\,\lambda_0\, \tilde\omega_{i_{k}} \tilde\omega_{ref}}\right).
\end{equation}
Also going from (\ref{intm}) to (\ref{intm1}) we have ignored the modulus assuming that, the individual terms are positive. 
Finally we get,
\begin{align}
    \begin{split}\label{finalCA2}
         C_{\kappa=1}^{(2)}
         =&\frac{V}{\delta^{d-1}}\frac{(\lambda_R\, \delta^{4-d})^{\mu} \delta^{-\nu}\, V^{\frac{\nu}{d-1}}}{2}\Bigg\{\sum_{k=1}^{d-1}\sum_{i_k=0}^{N-1}\log\left(\frac{b_{i_k}\,\lambda_R\,  \delta^{2-d}}{V^{\frac{1}{d-1}}\,\lambda_0\, m_R\, \tilde\omega_{ref}}\right)
         \\&-\frac{1}{2}\sum_{k=1}^{d-1}\sum_{i_k=0}^{N-1}\log\left(1+\frac{4}{(m_R\,\delta)^2}\sum_{k=1}^{d-1}\sin^2\left(\frac{\pi i_k}{N}\right)\right)\Bigg\}
    \end{split}
\end{align}\\
where, $\lambda_0$ can be set to either one of the expressions given in (\ref{zerol}).
The first sum above yields a factor of $\frac{V}{\delta^{d-1}}$ and  it is being multiplied  effectively by a factor $\log(\delta).$(upto some suitable factors to make it dimensionless inside the logarithm.) We had   already dealt with the second sum in the previous section, as it has appeared in $\lambda^{(1)}_{i}$ contribution. We will just use the expression given in (\ref{gendsum}) for that. Now if we focus on the first sum in (\ref{finalCA2}), we see that after performing the sum it gives a $N^{d-1}$ factor. The second sum in (\ref{finalCA2}) always grows as $N^{d-1}$ as evident from (\ref{gendsum}). We demand that the leading volume dependence coming from the first sum in (\ref{finalCA2}) will utmost be of the same order as free theory i.e $V.$  If assume this we can now choose the following,
\be \label{choice1}
\mu=1, \quad  \nu=(1-d).
 \ee
 Alternatively we can demand that the $\delta$ dependence can be $\delta^{6-2d}$ like the $\mathcal{O}(\lambda_R)$ contribution of $\mathcal{C}^{(1)}_{\kappa=1}.$ Then we could have chosen,
 \be \label{choice2}
\mu=1, \quad \nu=- \, d.
 \ee
 We can generalize this argument for the penalty factor order by order in higher order in $\lambda_R$ and we put some more details in the Appendix~(D).\\

\subsection{Flow equations}
 Here we would like to consider the flow equations for $\Delta {\mathcal C}_{\kappa=1}$ defined via
\be
\widetilde{\Delta {\mathcal C}}\equiv (\mathcal{C}_{\kappa=1}-\mathcal{C}_{\kappa=1}|_{\lambda_R=0})\frac{\delta^{d-1}}{V}\,.
\ee
Since, $\frac{V}{\delta^{d-1}}=N^{d-1}$, $\widetilde{\Delta {\mathcal C}}$ can be thought of as the complexity per degree of freedom. Now we want to consider the transformations \cite{subirbook}
$$
 \lambda_R \rightarrow b^{d-4} \lambda_R' , \quad\delta\rightarrow b\, \delta\,,
$$
with an infinitesimal change in $b$, namely $b=1+db$ which leads to $\lambda_R'=\lambda_R+d\lambda_R$.
Then straightforwardly (for $\mu=1, \nu= -d$ for the second block, so that it is subleading for large $V$), we find that up to linear order in $\lambda_R$,

\be
\frac{d\widetilde{\Delta{\mathcal C}}}{db}=2(4-d)\widetilde{\Delta {\mathcal C}} \,.
\ee
These equations are similar to the flow equations for $\lambda_R$ namely $\frac{d\lambda_R}{db}=(4-d)\lambda_R+O(\lambda_R^2)$ (which has also been used in deriving the form above). For $d>4$ we conclude that the flow for $\widetilde{\Delta {\mathcal C}}$ should be back towards $\widetilde{\Delta {\mathcal C}}=0,$ which is what we get when we turn off the coupling. The opposite happens for $d<4$. Quite pleasingly, this conclusion agrees with the RG picture (see e.g., \cite{subirbook}). 

\subsection{Comparison with Holography}
Here we briefly make a qualitative comparison with the holographic results.  From (\ref{d=4}) it is evident that the free theory part is always proportional to spatial volume ($V$). This is same as that of the expression for complexity coming from holographic calculation. But interestingly from (\ref{d=4}) we note that the $\mathcal{O}(\hat \lambda)$ terms are proportional to fractional power of the spatial volume (except for $d=2$ where there is no $V$ dependence in the $O(\hat\lambda)$ correction). Current holographic proposals will never produce such terms. If we consider the complexity equals volume conjecture, then using the results of \cite{Hung} we can easily see that terms with fractional power of volume will never occur. Complexity equals action gives rise to a logarithmic enhancement of the volume divergence but  still it doesn't give rise to fractional volume \cite{Hol6}. Also the occurrence of this fractional power is independent of the choice of the penalty factor (\ref{it}) as this feature shows up in the Gaussian part itself irrespective of whether we are suppressing the contribution from the second block or not.  Moreover from  the holographic side \cite{Alish, RathU}, the corrections due to a relevant deformation using existing holographic proposals have been considered leading to $V$ dependence in the complexity expression. Similar type of results for the complexity for $d=2$ has been derived using path-integral approach by considering relevant and marginal deformation of conformal field theory in \cite{Cap3}. In light of that, we should emphasize here that,  in the path-integral approach   (also in the context of holography) one starts from a conformal field theory and then considers perturbation around that. Now we will have a conformal field theory only  at the fixed points. In our setup we  can only access these fixed point perturbatively, i.e., in the epsilon expansion, hence the comparison has to  be made with great care.  

\section{Using circuit complexity for inferring fixed points}

In this section we will briefly generalize the discussion above for $\mathcal{C}_{\mathcal{N}\, \kappa}^{(1)}$ for a theory with ${\mathcal N}$ scalars. We want to study this theory to see if complexity arguments can be used to infer the existence of fixed points.  We will consider the following Hamiltonian:
\begin{equation}\label{newham}
    H=\frac{1}{2}\int d^{d-1}x \sum_{a=1}^{\mathcal{N}}\Big[\pi_a(x)^2+\vec \nabla \phi_a(x)^2+m^2\phi_a(x)^2+\frac{\hat \lambda_1}{12}\Big(\phi_a(x) \phi_a(x)\Big)^2+ \frac{\hat\lambda_2}{12}\phi_{a}^4(x)\Big]\,.
\end{equation}
Here the $\hat \lambda_1$ term has $O(N)$ symmetry while the $\hat\lambda_2$ term breaks this symmetry. There are interesting fixed points that this theory allows for, which have been studied in the epsilon expansion \cite{phi4book}.\par
Next we solve  the ground state wavefunction perturbatively  in $\hat\lambda_1$ and $\hat\lambda_2$ as before. Then considering only the contribution from the first block and concentrating  first on $\hat \lambda_1$ correction, we get the following eigenvalues,

\begin{align}
\begin{split}\label{evalN}
    \lambda^{(1)}_{a\, i_{k}} = \frac{\tilde\omega_{a\, i_{k}}}{\delta} + \Big\{\frac{3\lambda_1}{2 N}\Big(\frac{2}{\tilde\omega_{a\, i_{k}}(\tilde\omega_{a\, i_{k}}+\tilde\omega_{a\,(N-i_{k})})}\Big)\Big\}
    +\sum_{b=a+1}^{\mathcal{N}}\Big\{\frac{\lambda_1}{ N}\Big(\frac{1}{\tilde\omega_{a\, i_k}(\tilde\omega_{a\, i_{k}}+\tilde\omega_{b\,(N-i_{k})})}\Big)\Big\},\\&\quad 
\end{split}
\end{align}
where for technical simplification we are considering $N$ odd only.
Here $a$ index run from $1$ to $\mathcal{N}$ counting the number of components of vector field and each of these $i_{k}$ as usual runs from $0$ to $N-1$ for every $a.$ Basically now we have $\mathcal{N}$ number of $\lambda^{(1)}_{i_{k}}$ eigenvalues compared to the $\mathcal{N}=1$ case.  We also define $\lambda_1=\frac{\hat \lambda_1\,\delta^{-d}}{24}$.
The first interaction term above i.e., the $b$ independent term, same as the eigenvalues for  the  $\lambda_1 \phi^4$ theory, comes from the $\lambda_1 X_a(\vec{n})^4$ term in the Hamiltonian, while the second set of terms involving sum over the  $b$ index arise from the $2\, \lambda_1 X_a(\vec{n})^2X_b(\vec{n})^2$ terms in the Hamiltonian.
Correspondingly, the complexity $\mathcal{C}_{\mathcal{N}\, \kappa}^{(1)}$  is given by,
\begin{align}
\begin{split}\label{cN}
\mathcal{C}_{\mathcal{N}\, \kappa}^{(1)}=&\frac{1}{2^\kappa}\sum_{a=1}^{\mathcal{N}}\sum_{k=1}^{d-1}\sum_{i_{k}=0}^{N-1}\Big(\Big|\log\frac{\tilde \omega_{a\, i_{k}}}{\tilde \omega_{ref}}\Big|^{\kappa}+\frac{3\lambda_1\,\kappa\, \delta}{2\,N}\Big\{\frac{1}{ \tilde \omega_{a\, i_{k}}^3}+\frac{2}{3}\,\sum_{b=a+1}^{\mathcal{N}}\Big(\frac{1}{ \tilde\omega^2_{a\, i_{k}}(\tilde\omega_{a\, i_{k}}+\tilde\omega_{b\, i_{k}})}\Big) \Big\}\Big|\log(\frac{\tilde \omega_{a\, i_{k}}}{\tilde \omega_{ref}})\Big|^{\kappa-1}\Big)\\&+\mathcal{O}(\lambda_1^2) .
\end{split}
\end{align}
For the case of $\mathcal{N}=1$ the $b$ sum does not contribute and the $a$ sum contributes one term, which leads to the same complexity as the $\lambda \, \phi^4$ case. 
Then finally from  \eqref{cN} we get, 
\begin{align}
\begin{split}\label{cN1}
\mathcal{C}_{\mathcal{N}\, \kappa}^{(1)}
=  \mathcal{N}\mathcal{C}_{free}+\hat \lambda_1\, \frac{(\mathcal{N}+1)(\mathcal{N}+2)}{6}\mathcal{C}_{int}  
\end{split}
\end{align}\\
where $\mathcal{C}_{free}$ is the complexity for the free theory and $\mathcal{C}_{int}$ is the interaction term sans the coupling constant for the $\lambda\, \phi^4$ theory corresponding to $\mathcal{N}=1$.

The additional $\frac{\hat\lambda_2\phi_a^4}{24}$ interaction piece of the Hamiltonian contributes the following additional terms to the eigenvalues \eqref{evalN}.
\begin{align}
\begin{split}
&\lambda^{(1)}_{a\, i_{k}} \rightarrow \lambda^{(1)}_{a\, i_{k}}+ \Big\{\frac{3\lambda_2}{2 N}\Big(\frac{2}{\tilde\omega_{a\, i_{k}}(\tilde\omega_{a\, i_{k}}+\tilde\omega_{a(N-i_{k})})}+\frac{2}{\tilde\omega_{a\, i_{k}}(\tilde\omega_{a\, i_{k}}+ \tilde\omega_{a\,(\frac{N}{2}-i_{k})})}\Big)\Big\},\quad  N:even\\
    &\lambda^{(1)}_{a\, i_{k}} \rightarrow \lambda^{(1)}_{a\, i_{k}}  + \Big\{\frac{3\lambda_2}{2 N}\Big(\frac{2}{\tilde\omega_{a\, i_{k}}(\tilde\omega_{a\, i_{k}}+\tilde\omega_{a\,(N-i_{k})})}\Big)\Big\},\quad  N:odd
\end{split}
\end{align}
where $\lambda_2=\frac{\hat\lambda_2\delta^{-d}}{24}.$
We then have the following expression for complexity of the theory given by \eqref{newham},
\begin{align}\label{cubica}
\begin{split}
\mathcal{C}_{\mathcal{N}\, \kappa}^{(1)}
=&\mathcal{N}\mathcal{C}_{free}+\Big(\hat \lambda_1\, \frac{(\mathcal{N}+1)(\mathcal{N}+2)}{6}+\hat \lambda_2\, \mathcal{N}\Big)\mathcal{C}_{int}.  
\end{split}
\end{align}
\subsection*{Connection with RG}
Now from (\ref{cubica}), we see the following. First in the subsequent discussion will use renormalized coupling $\{\lambda_{R_{1}}, \lambda_{R_{2}}\}$ instead of $\{\hat \lambda_1,\hat \lambda_2\}.$  When we express the bare mass contribution from the free part in terms of the renormalized mass we will get a term in the complexity that is proportional to $\displaystyle\frac{V}{\delta^{d-1}}(\lambda_{R_1}\frac{\mathcal{N}+2}{3}+\lambda_{R_2})\delta^{4-d}$. Furthermore, the coupling constants in the interaction part appear in the combination $(\lambda_{R_1}\frac{(\mathcal{N}+2)(\mathcal{N}+1)}{6}+\mathcal{N}\lambda_{R_2})$. Since the weights of the $ \lambda_{R_{1}}, \lambda_{R_{2}}$ are different, it is clear that when $\mathcal{N}$ is large, there will be a larger contribution\footnote{For this discussion, imagine $\lambda_{R_{1}}\sim \lambda_{R_{2}}$ to be of similar magnitude.} from $\lambda_{R_{1}}$ compared to $\lambda_{R_{2}}$. Then it is clear that there must be some intermediate $\mathcal{N}=\mathcal{N}_c$ when there is a crossover between which term dominates.  In fact in the epsilon expansion the beta functions admit the following fixed points at leading order \cite{phi4book}
\be
(\lambda_{R_{1}*},\lambda_{R_{2}*})=(0, \frac{\epsilon}{3}), (\frac{\epsilon}{\mathcal{N}}, \frac{\epsilon(\mathcal{N}-4)}{3 \mathcal{N}}), (\frac{3\epsilon}{8+\mathcal{N}},0), (0,0)\,.
\ee
The last corresponds to the Gaussian fixed point. The first corresponds to the Ising fixed point (decoupled theory of $\mathcal{N}$ scalars), the third to the Heisenberg/$O(\mathcal{N})$ fixed point. The second corresponds to the cubic anisotropic fixed point where {\it both} couplings are turned on. Note here that in the $\mathcal{N}$ large limit, $\lambda_{R_{1}*}\rightarrow 0$. This is also to be expected from the circuit complexity expression where it is less favourable to turn $\lambda_{R_{1}}$ on in this limit. 
From the complexity expression proportional to $\mathcal{C}_{free},$ for $\lambda_{R_{1}}=\lambda_{R_{2}} \frac{3}{(\mathcal{N}+2)}$,  and for $\mathcal{C}_{int},$ for $\lambda_{R_{1}}=\lambda_{R_{2}} \frac{6 \mathcal{N}}{(\mathcal{N}+1)(\mathcal{N}+2)}$ in the plane of couplings, it is ``equally favourable'' to turn on both couplings. Hence it could be expected that one is allowed to perturbatively turn on both couplings in order to go away from the Gaussian, Ising and $O(\mathcal{N})$ fixed points. While this stops short of proving the existence of the cubic anistotropic fixed point using the complexity perspective, it hints at its existence (of course for $d<4$ since perturbative circuit complexity breaks down for $d>4$ even here).

\section{An alternative construction}
Here we briefly discuss an alternative construction of the circuit, and we will also estimate the complexity\footnote{This is also being considered in \cite{Cotler4}.}.   The advantage of the considerations so far is that we were able to give a meaningful geometrical interpretation to the complexity calculation. We were able to provide an optimal circuit (minimizing  $\mathcal{C}_{\kappa=1}$ functional)  which represents the wavefunction upto $\mathcal{O}(\hat \lambda).$ A by product of our calculation was that we were able to minimize the circuit depth and get an answer for the complexity. The downside of this calculation was that there are several ambiguities that enter  and we had to make a choice for the penalty factors to make the $\hat \lambda \rightarrow 0$ limit well defined. Now we want to present an alternate calculation which will help us to avoid the problems associated with the non-uniqueness  of the $\mathcal{O}(\hat \lambda)$ block.  The flip-side  of this alternative approach is that we do not know how to give it a geometrical interpretation, leaving this issue for future work.   Below we discuss our construction for $N=2$ case for $d=2$.\par  We start with Gaussian reference state by setting $\lambda_0=0$ in  (\ref{refstate}). We consider first the scaling operators 
\be
\tilde O_{1}=\frac{\tilde x_0.\tilde p_0+\tilde p_0.\tilde x_0}{2}= \tilde x_0.\tilde p_0-\frac{i}{2},\,\,\tilde O_2=\tilde x_1.\tilde p_1-\frac{i}{2},
\ee
and act on the reference state with them. This gives

 \begin{align}
 \begin{split} \label{int}
& \exp(i\,\epsilon\, \alpha_2\tilde O_{2}) \exp(i\, \epsilon\, \alpha_1\, \tilde O_{1})\psi^{s=0}(\tilde x_0,\tilde x_1)=\\&\mathcal{N}^{s=0}\, \exp\Big[-\frac{\tilde \omega_{ref}}{2}\Big((\exp(\epsilon\alpha_1)\tilde x_0)^2+(\exp(\epsilon \alpha_2)\tilde x_1)^2\Big].
 \end{split}
 \end{align}
 Then we can set,
 \begin{align}
 \begin{split}
 & \alpha_{1}=\frac{1}{2\,\epsilon}\log\Big(\frac{a_1}{\tilde \omega_{ref}}\Big)+\mathcal{O}(\hat \lambda^2),\\&
 \alpha_{2}=\frac{1}{2\,\epsilon}\log\Big(\frac{a_2}{\tilde \omega_{ref}}\Big)+\mathcal{O}(\hat \lambda^2).
  \end{split}
 \end{align}
 
 Here $\epsilon$ is a small parameter and $a_1, a_2$ are defined in (\ref{eq3.18}). 
 Then we consider the following operators,
\be \label{new}
\tilde O_3=\hat \lambda\,\Big( \frac{\tilde x_0^3\, \tilde p_{0}+\tilde p_{0}\, \tilde x_{0}^3}{2}\Big)=\hat \lambda\,  \Big(\tilde x_0^3\,\tilde p_0-\frac{3\,i\, \tilde x_{0}^2}{2}\Big),\,\,\tilde O_4=\hat \lambda\,\Big(\tilde x_1^3\,\tilde p_1-\frac{3\,i\, \tilde x_{1}^2}{2}\Big).
\ee
These two operators will be responsible for fixing the coefficients of $\tilde x^4,$ so their coefficient will be of the $\mathcal{O}(\hat \lambda^0).$  We will use this fact to approximate our calculation. We use the following unitary,
\be
\exp\Big(i\,\epsilon \, \alpha_3\,\frac{p_0}{x_0} \tilde O_3\Big)\approx 1+i\, \epsilon\, \alpha_3\,\frac{p_0}{x_0}\, \tilde O_3+\mathcal{O}(\hat \lambda^2).
\ee
Similarly, for $\tilde O_4$ we have the corresponding coefficient $\alpha_4.$ $p_0$ and $x_0$ are arbitrary constants with appropriate dimensions and the ratio $\frac{p_0}{x_0}$ is positive definite.
  Here,
  \be
   \alpha_3=\frac{x_0}{\epsilon\, p_0\,\hat\lambda}\Big(\frac{a_3}{2\, a_1}\Big)+\mathcal{O}(\hat \lambda),\quad \alpha_4=\frac{x_{0}}{\epsilon\, p_0\,\hat\lambda}\Big(\frac{a_4}{2\, a_2}\Big)+\mathcal{O}(\hat \lambda).
  \ee
 Now we consider the following two operators,
 \begin{align}
 \begin{split} \label{new1}
\tilde O_{5}=\hat \lambda\Big(\frac{\tilde x_1^2 \, \tilde x_0\,\tilde p_0+\tilde p_0\,\tilde x_1^2 \, \tilde x_0}{2}\Big)=\hat \lambda\,\tilde x_1^2\Big(\tilde x_0\,\tilde p_0-\frac{i}{2}\Big), \quad \tilde O_{6}=\hat \lambda\, \tilde x_0^2\Big(\tilde x_1\, \tilde p_1-\frac{i}{2}\Big).
 \end{split}
 \end{align}
 Then we act on the wavefunction with these two operators: $1+i\,\epsilon\, \alpha_5\, \frac{p_0}{x_0}\, \tilde O_5$ and $1+i\,\epsilon\, \alpha_6\,\frac{p_0}{x_0}\, \tilde O_6$   and try to fix $\alpha_5,\alpha_6$.
However we can determine only one of them at this moment. We determine $\alpha_5$ without loss of any generality, but we assume here that $\alpha_6$ is of the order $\mathcal(\hat \lambda^0),$ although we  cannot uniquely determine it at this stage. 
\be
\alpha_5=\frac{1}{\hat \lambda} \Big(\frac{a_5}{2\, a_1\, \epsilon}\frac{x_0}{p_0}\Big)-\frac{a_2}{2\,a_1}\, \alpha_6+\mathcal{O}(\hat \lambda).
\ee
 We notice that we have generated two extra terms, $\tilde x_0^2$ and $\tilde x_1^2$ with $\mathcal{O}(\hat\lambda)$ coefficients.  Then we act on them again with two unitaries constructed from  $\tilde O_1$ and $\tilde O_2,$    $$1+i\,\epsilon\, \alpha_7 \,\tilde O_1\quad 1+i\,\epsilon\, \alpha_8\,\tilde O_2.$$ We find,
 \be
 \alpha_7=\frac{3}{2\,\epsilon}\Big(\frac{a_3}{2\, a_1^2}\Big)+\hat \lambda\, \frac{p_0}{x_0}\frac{\alpha_6}{2\,a_1}+\mathcal{O}(\hat \lambda^2), \alpha_8=\frac{3}{2\,\epsilon}\Big(\frac{a_4}{2\, a_2^2}\Big)+\frac{1}{4\,\epsilon}\frac{a_5}{a_1\,a_2}- \hat \lambda\, \frac{p_0}{x_0}\frac{\alpha_6}{2\,a_1}+\mathcal{O}(\hat \lambda^2).
 \ee
 After doing this  we reproduce the target state exactly (\ref{eq3.17}).  $a_3, a_4, a_5$ are defined in (\ref{eq3.18}). $\alpha_7$ and $\alpha_8$ are of the $\mathcal{O}(\hat \lambda).$ We get after reinstating appropriate factor of $\delta$,
 \begin{align}
 \begin{split}
 \mathcal{D}(U)&=\,\sum_{i=1}^{8}|\alpha_i|\\&=\frac{1}{2\,\epsilon}\Big[\log\Big|\frac{\tilde \omega_{0}}{\tilde \omega_{ref}}\Big|+\log\Big|\frac{\tilde \omega_{1}}{\tilde \omega_{ref}}\Big|+\frac{\hat\lambda}{32\, \delta}\Big(\frac{3 \tilde \omega_{0}+\tilde \omega _{1}}{\tilde \omega _{0}^3 (\tilde \omega_{0}+\tilde \omega _{1})}+\frac{\tilde \omega_{0}+3 \tilde \omega_{1}}{\tilde \omega_{1}^3 (\tilde \omega_{0}+\tilde \omega_{1})}\Big)\Big]\\&+\frac{x_0}{\epsilon\,  p_0}\,\Big(\frac{(\tilde \omega_0^2+\tilde\omega_1^2) }{192\,\delta^2 \tilde \omega_0^2\tilde \omega_1^2}+\frac{ 1}{16\,\delta^2\, (\tilde \omega _1+\tilde \omega _0)\tilde \omega_0}\Big)+\hat \lambda\, \Big(1-\frac{\tilde \omega_1}{\tilde \omega_0}\Big)\, \alpha_6\\&+\frac{3}{2\,\epsilon}\Big(\frac{\hat \lambda(\tilde \omega_0^3+\tilde \omega_1^3) }{192\,\delta\,  \tilde \omega_0^3\,\tilde \omega_1^3}+\frac{ \hat \lambda }{48\,\delta\,(\tilde \omega_1+\tilde \omega_0)\tilde \omega_0\tilde \omega_1}\Big). \end{split}
 \end{align}
 
 As $\Big(1-\frac{\tilde \omega_1}{\tilde \omega_0}\Big) >0 ,$ when $\alpha_6=0$ this will be minimized giving   \begin{align}
 \begin{split}\label{circuitdepth}
 \mathcal{D}(U)&=\frac{1}{2\,\epsilon}\Big[\log\Big|\frac{\tilde \omega_{0}}{\tilde \omega_{ref}}\Big|+\log\Big|\frac{\tilde \omega_{1}}{\tilde \omega_{ref}}\Big|+\frac{ \hat \lambda}{32\, \delta}\Big(\frac{1}{\tilde \omega_0^3}+\frac{1}{\tilde \omega_1^3}+ \frac{2}{\tilde \omega _{0}^2 (\tilde \omega_{0}+\tilde \omega _{1})}+\frac{2}{\tilde \omega_{1}^2 (\tilde \omega_{0}+\tilde \omega_{1})}\Big)\\&+\frac{x_0}{ p_0}\,\Big(\frac{(\tilde \omega_0^2+\tilde\omega_1^2) }{96\,\delta^2 \tilde \omega_0^2\tilde \omega_1^2}+\frac{1}{8\,\delta^2\, (\tilde \omega _1+\tilde \omega _0)\tilde \omega_0}\Big)+\Big(\frac{\hat \lambda(\tilde \omega_0^3+\tilde \omega_1^3) }{64\,\delta\,  \tilde \omega_0^3\,\tilde \omega_1^3}+\frac{\hat  \lambda }{16\,\delta\,(\tilde \omega_1+\tilde \omega_0)\tilde \omega_0\tilde \omega_1}\Big)\Big]. \end{split}
 \end{align}
Now we make several comments.
\begin{itemize}
\item We note that the first line of (\ref{circuitdepth}) is same as that of $\mathcal{C}_{\kappa=1}$ in (\ref{eq4.13}) for $N=2$ and $d=2.$ This comes from the Gaussian part. This can be readily extended for arbitrary number of oscillator ($N$) and the conclusion that we drawn regarding the RG flow in the previous section will remain same. 
\item The first two terms in the second line of (\ref{circuitdepth}) have no counterpart in free theory. So we have to introduce a penalty factors for $\alpha_3,\alpha_4,\alpha_5,$ 
\be
\mathcal{D}(U)=\sum_{i=1}^{8}p_{i}|\alpha^{i}|,
\ee
with, $p_1=p_2=p_7=p_8=1$ and $p_3, p_4 , p_5$ proportional $\hat \lambda$.\footnote{ In large $N$ limit the $\mathcal{O}(\hat \lambda)$ part will be again proportional to $\frac{1}{\omega_i^3}$ as before.}
Then (\ref{circuitdepth}) will have smooth $\hat \lambda \rightarrow 0$ limit.
\item We notice that this same set of operators has been used in the proposal of interacting cMERA for $\phi^4$ theory recently \cite{Cotler2,Cotler3}.  Working perturbatively in $\mathcal{O}(\hat \lambda)$ we can work out the algebra satisfied by these operators.   \begin{align}
 \begin{split} \label{alg}
&[\tilde O_1,\tilde O_2]=0,[\tilde O_3,\tilde O_4]=0,[\tilde O_1 ,\tilde O_3]=-2\,i\, \tilde O_3,
[\tilde O_1,\tilde O_4]=0,\\& [\tilde O_2,\tilde O_3]=0,[\tilde O_2,\tilde O_4]=-2\,i\, \tilde O_4,
[\tilde O_1,\tilde O_5]=0,[\tilde O_2,\tilde O_5]=-2\,i\,\tilde O_5,\\&[\tilde O_3,\tilde O_5]=\mathcal{O}(\hat \lambda^2),
 [\tilde O_4, \tilde O_5]=\mathcal{O}(\hat \lambda^2).
\end{split}
\end{align}      
From these one can easily see that these operators can have at least  4 dimensional representations. However  given a Gaussian reference state currently we are not able to utilize this algebra  to geometrize this problem  which would have helped us to achieve \textit{``minimal"} circuit depth.  Moreover, we can use these matrix representations coming from (\ref{alg})  and try to take a linear combination of them so that we can reproduce $\tilde M $ given in (\ref{lincomb}) coming from the geodesic analysis.  We can see that there are no such linear combinations that allow us to reproduce $\tilde M$ given in (\ref{lincomb}).  \item    It would be nice from the tensor network point of view  to test whether the cMERA proposal in \cite{Cotler2,Cotler3} based on these  gates achieve minimal complexity or not.  Since the Gaussian part of the expression will be the same as our earlier analysis, we expect that the connection we have found with the RG perspective will continue to hold. The interesting question here is, if it will allow us a more refined understanding, in the sense of being able to detect fixed points.
\item There is another interesting possibility. Instead of defining $\tilde O_3,\tilde O_4,\tilde O_5$ with $\hat \lambda$ as in (\ref{new}) and (\ref{new1}) we could have pushed this $\hat \lambda$ inside $\alpha_3,\alpha_4$ and $\alpha_5.$ Then we will have, 
\begin{align}
 \begin{split}\label{circuitdepth1}
 \mathcal{D}(U)&=\frac{1}{2\,\epsilon}\Big[\log\Big|\frac{\tilde \omega_{0}}{\tilde \omega_{ref}}\Big|+\log\Big|\frac{\tilde \omega_{1}}{\tilde \omega_{ref}}\Big|+\frac{ \hat \lambda}{32\, \delta}\Big(\frac{1}{\tilde \omega_0^3}+\frac{1}{\tilde \omega_1^3}+ \frac{2}{\tilde \omega _{0}^2 (\tilde \omega_{0}+\tilde \omega _{1})}+\frac{2}{\tilde \omega_{1}^2 (\tilde \omega_{0}+\tilde \omega_{1})}\Big)\\&+\hat \lambda\, \frac{x_0}{ p_0}\,\Big(\frac{(\tilde \omega_0^2+\tilde\omega_1^2) }{96\,\delta^2 \tilde \omega_0^2\tilde \omega_1^2}+\frac{1}{8\,\delta^2\, (\tilde \omega _1+\tilde \omega _0)\tilde \omega_0}\Big)+\Big(\frac{\hat \lambda(\tilde \omega_0^3+\tilde \omega_1^3) }{64\,\delta\,  \tilde \omega_0^3\,\tilde \omega_1^3}+\frac{\hat  \lambda }{16\,\delta\,(\tilde \omega_1+\tilde \omega_0)\tilde \omega_0\tilde \omega_1}\Big)\Big]. \end{split}
 \end{align}
 Then $\mathcal{D}(U)$ will have smooth $\hat \lambda \rightarrow 0$ limit without having to choose penalty factors proportional to $\mathcal{O}(\hat \lambda)$. But in this case unlike what is shown in (\ref{alg}), the algebra between these operators does not close. We leave these issues for future investigations.
\end{itemize}

  \section{Discussions}
  We have shown that there is a connection between circuit complexity and renormalization group flows. Namely, we found using the scaling equations for circuit complexity that, for $d>4$ the perturbative calculation for circuit complexity breaks down. Put differently, it becomes unfavourable to turn on a $\phi^4$ coupling for $d>4$. This conforms with the RG picture, that it is the Gaussian fixed point that is the stable fixed point for $d>4$. We also saw that it is possible to argue the existence of the cubic anisotropic fixed point for $\mathcal{N}$ scalars. There are several interesting questions to explore:
 \begin{itemize}
 \item One of the interesting conclusions from our calculations was that turning on interactions lead to fractional dependence on the volume. While in the large volume limit, these would be subleading, it still warrants  the question how holographic calculations would see such fractional dependence. This conclusion is unambiguous and is unchanged by what is happening to the ``ambiguous'' block.
 \item 
 Our construction needed us to fix certain ambiguities. In order to make the circuit complexity calculation perturbative, we were forced to make some choices which translated into making the cost functional depend on the coupling. This on its own is not an essential drawback in what we have done, since even in holographic entanglement entropy calculations, the entropy functional depends on the coupling. Nevertheless, we do not have a first principle way of fixing these ambiguities we encountered. 
 \item 
 Our final circuit complexity formulas involved an integral transform. While this is true for the free theory and leading order in perturbation, it is easy to see that it must be true at higher orders in perturbation as well, since the lattice sums can be handled similarly. Especially after picking up the correct $s$-residue in eq.(\ref{A6}) we are left with a Laplace transform over the Schwinger parameter $t$. A question that arises is if there is a physical interpretation that can be given to this variable. 
 \item
 In section 5, we showed that there is an alternative set of gates which could be used for circuit complexity calculation which will enable us to use a Gaussian reference state as opposed to a nearly Gaussian reference state we have used in this paper. It will be interesting to develop the geometric picture further for these set of gates. It will also be interesting to develop the ideas in this paper for other interacting theories involving fermions and gauge fields. The idea would be to use circuit complexity to say which theories are reasonable theories from a quantum computer's point of view--any circuit that is prohibitive in allowing a calculation would not be allowed for instance. After all, eventually we would like to understand what is so special about the standard model from this view point.
 \end{itemize}

  \subsection*{Acknowledgments} 
  We thank Michal P. Heller, Ro Jefferson, Chethan Krishnan, Esperanza Lopez, Jose Juan Melgarejo, Pratik Nandy, Apoorva Patel, Barry Sanders, Tadashi Takayanagi, Javier Molina Vilaplana  and especially Rob Myers for useful discussions.  We thank Rob Myers for useful comments on this manuscript. AB also thanks S.~Prem Kumar of Department of Physics, University of Swansea, Marika Taylor of University of Southampton,  Physics Department, Department of Physics, University of Murcia, Karl Landsteiner of IFT Madrid,  Galileo Galilee  Institute, Florence, Julian Sonner of University of Geneva and Michal Heller of Max Planck Institute for Gravitational Physics, Potsdam for generously hosting him during  the course of this work.   AB is supported by JSPS Grant-in-Aid for JSPS fellows (17F17023). A.S. acknowledges support from a DST Swarnajayanti Fellowship Award DST/SJF/PSA-01/2013-14.  
      
       \begin{appendix}
      \section{Matrix representation  of Unitary operators}
     As discussed in the main text, we here give the explicit matrix  representations  of the unitary operators which are used to construct the circuit. 
   \begin{align}
   \begin{split} \nonumber
& M_{11}=diag(1,0,1,2,0)
,
 M_{22}=diag(0,1,1,0,2),
 M_{33}= diag(-1,-\frac{1}{4},-\frac{13}{4},-3,-\frac{1}{2}),
\end{split}
\end{align}
\begin{align}
   \begin{split} 
 &M_{44}=diag(-\frac{1}{4},-1,-\frac{13}{4},-\frac{1}{2},-3),
 M_{77}=diag(-\frac{13}{4},0,-\frac{13}{4},-\frac{33}{2},0), M_{88}=diag(0,-\frac{13}{4},-\frac{13}{4},0,-\frac{33}{2})\\&
M_{55}=\left(
\begin{array}{ccccc}
 -\frac{3}{2} & 0 & 0 & 0 & 0 \\
 0 & 0 & 0 & 0 & 0 \\
 0 & 0 & -6 & 0 & 0 \\
 0 & 0 & 0 & -5 & -1 \\
 0 & 0 & 0 & -5 & -1 \\
\end{array}
\right),
 M_{66}=\left(
\begin{array}{ccccc}
 0 & 0 & 0 & 0 & 0 \\
 0 & -\frac{3}{2} & 0 & 0 & 0 \\
 0 & 0 & -6 & 0 & 0 \\
 0 & 0 & 0 & -1 & -5 \\
 0 & 0 & 0 & -1 & -5 \\
\end{array}
\right).
\end{split}
\end{align}
We can show that in $U(s)=\exp\Big(\tilde M s\Big) $, $\tilde M$ is given by,
\begin{align}
\begin{split}
\tilde M=\alpha  M_{11}+\beta  M_{22}+\gamma  M_{33}+\delta  M_{44}+\zeta  M_{55}+\tau  M_{66} +\kappa  M_{77}+\mu M_{88},
\end{split}
\end{align}
where, 
\begin{align}
\begin{split}
&\alpha = \frac{1}{160} (472 a-108 b+32 c+33 d-164 \Delta a+136 (\Delta c+5 \mu )),\\&
\beta = \frac{1}{16} (88 a-4 b-24 c+5 d+84 \Delta a-24 \Delta c-68 \mu ),\\&
\gamma = -2 a-b+2 c+\frac{d}{4}-3 \Delta a+2 \Delta c+10 \mu,\quad
\delta =\frac{1}{2} (8 a-4 c+d+8 \Delta a-4 \Delta c-20 \mu ),\\&
\zeta = -\frac{d}{6},\quad
\tau = -\frac{d}{6},\quad
\kappa = \frac{1}{40} (24 a+4 b-16 c+d+12 \Delta a-8 (\Delta c+5 \mu )),\\&
a=y_1(1)-\rho_1(1),\quad \Delta a=2\rho_1(1), \quad b=y_2(1),\\&  c=y_3(1)-\rho_3(1)\sin(\theta_3(1)),\quad \Delta c=2\rho_3(1)\sin(\theta_3(1)),\quad d=\rho_3(1)\cos(\theta_3(1)).
\end{split}
\end{align}
Note that there is  certain amount of arbitrariness in identifying. We could have set either $\kappa$ or $\mu$ to zero. 

   \section{ Counting of dimension of the target matrix and the basis elements for arbitrary $N$:}

In this appendix, we discuss in detail the counting of the number of components in the basis.   From (\ref{eq4.5}) and (\ref{eq4.6}), it is evident that we have to allow for the quadratic term in the basis $\vec{v}.$ In general the basis will of the following form,
\be  \label{eqB.1}
\vec{v}=\Big\{\tilde x_0\cdots \tilde x_{N},\tilde x_0^2\cdots \tilde x_{N}^2, \tilde x_i\tilde x_j\Big\},
\ee
where $i \neq j$ and both $i$ and $j$ run from $0$ to $N-1.$
Now we give an explicit counting for the number of terms in the basis. 
For the odd $N$ case the minimal basis required to span the target state  will always be:\\ \\
$\{0, 1, 2, 3, ...., N - 1\} \rightarrow$ ``\textit{unambiguous}" part\\ 
$\{00, 01, 02, 03, ..., 0 (N - 1), 11, 12, 13, 14, .., 1 (N - 1), 22, 23, 24, .., 2 (N - 1), ... ... ... .., (N - 1) (N - 1)\} \rightarrow $ ``\textit{ambiguous}" part.

Here we have simply denoted $\tilde x_i$'s by $i$ and $\tilde x_i\tilde x_j$ by $i,j.$
The total dimension ($D$) of the matrix for the $N$ odd case:

\be
D=N + \frac{1}{2} N (N + 1)
\ee\\
For the even $N$ case the basis will include terms:\\ \\
$\{0, 1, 2, 3, ...., N - 1 \}\rightarrow$ ``\textit{unambiguous}" part\\ \\
For the ``\textit{ambiguous}" part:\\ \\
The set of basis will include the term $ab$ if either one of the conditions given below are met:
\begin{itemize}
\item $(a + b)$ is a multiple of 2 
\item $(a + b) = N/2$ or $(a + b) = 3 N/2$
\end{itemize}
The number of terms that satisfy these conditions equals $\Big(\frac{N(N + 2)}{4}\Big),$ if $N$ is divisible by 4 and $\Big(\frac{N}{2}+\frac{N(N + 2)}{4}\Big)$ otherwise\\ \\
Therefore the dimension ($D$) of the target matrix $A_T$ in the $N$ even case is:
 \be
 D=N+\Big(\frac{N(N + 2)}{4}\Big)
 \ee
if $N$ is divisible by 4
\be
D=N+\Big(\frac{N}{2}+\frac{N(N + 2)}{4}\Big)
\ee
if $N$ is not divisible by 4.
 
 To summarize, the number of basis elements for the  linear part of the basis is $N$ and the number of quadratic elements in the basis grows as $N^2.$  As far as we can see, this is the minimal  way of extending the basis to include the quadratic terms and this is sufficient to produce all the terms in the wavefunction upto $\mathcal{O}(\lambda)$.

       \section{A general method to do the lattice sums}
       We are dealing with lattice (multiple) sums involving powers and logarithms of $m^2+4 \Omega^2\sum_{k=1}^{d-1}\sin^2\left(\frac{\pi i_k}{N}\right)$. Similar sums occur regularly in lattice field theory. The technique we will focus on is similar to what can be found in \cite{phi4book}, except that we will generalize this approach to handle any power and logarithms. This is best handled by making use of an integral transform and writing
       \be\label{A1}
       \int_{\mathcal{C}} ds\, f(s) (y+x)^s\,,
       \ee
     where for evaluating $\log(y+x)$ we choose $f(s)=1/s^2$ and for $(y+x)^a$ we choose $f(s)=1/(s-a)$. The contour ${\mathcal{C}}$ encircles the pole.  Then we make use of the Schwinger parametrization to write
     \be
     (y+x)^s=\frac{1}{\Gamma(-s)}\int_{0}^\infty dt\, e^{-t(y+x)} t^{-s-1}\,.
     \ee
      For our sums $x= 4 \Omega^2\sum_{k=1}^{d-1}\sin^2\left(\frac{\pi i_k}{N}\right)$ and $y=m^2$. Using\footnote{this is for a particular $i_k$.}
      \be
      \sum_{i_k=0}^{N-1}\sin^{2\alpha}\left(\frac{\pi i_k}{N}\right)=N \frac{(\frac{3}{2})_{\alpha-1}}{2(2)_{\alpha-1}}\,,
      \ee
      where $(a)_b=\Gamma(a+b)/\Gamma(a)$ is the Pochhammer symbol, we find
      \be
      \sum_{k=1}^{d-1} \sum_{i_k=0}^{N-1} e^{-t 4 \Omega^2 \sum_{k=1}^{d-1}\sin^2\left(\frac{\pi i_k}{N}\right)}=N^{d-1}\left(e^{-2t\Omega^2}I_0(2t \Omega^2)\right)^{d-1}\,,
       \ee
       where $I_0$ is a modified Bessel function of the first kind. Notice that in the form of the RHS, we can choose $d$ to be fractional as well! Thus we now have a way to compute complexity in the epsilon expansion, say $d=4-\epsilon$. So in all we now have
      
        \be
  \sum_{k=1}^{d-1}\sum_{i_k=0}^{N-1}   (m^2+4 \Omega^2\sum_{k=1}^{d-1}\sin^2\left(\frac{\pi i_k}{N}\right))^s=N^{d-1}\frac{1}{\Gamma(-s)}\int_{0}^\infty dt\, e^{-t m^2} t^{-s-1}\left(e^{-2t\Omega^2}I_0(2t \Omega^2)\right)^{d-1}\,.
     \ee
     We now move to a dimensionless variable using $t\rightarrow t/\Omega^2,$ which gives 
        \be\label{A6}
 \sum_{k=1}^{d-1} \sum_{i_k=0}^{N-1}   (m^2+4 \Omega^2\sum_{k=1}^{d-1}\sin^2\left(\frac{\pi i_k}{N}\right))^s=N^{d-1}\Omega^{2s}\frac{1}{\Gamma(-s)}\int_{0}^\infty dt\, e^{-t \hat m^2} t^{-s-1}\left(e^{-2t}I_0(2t)\right)^{d-1}\,.
     \ee
   Here $\hat m=m \delta$. For $d=2,3$,  the $t$-integral at this stage can be done on  mathematica yielding hypergeometric functions. Explicitly, for $d=2$ we get the RHS of eq.(\ref{A6}) to be
   \be
   N\Omega^{2s} (2+\hat m^2)^s {}_2F_1[\frac{1-s}{2},-\frac{s}{2},1;\frac{4}{(\hat m^2+2)^2}]\,,
   \ee
   and for $d=3$ we have
   \be
   N^2\Omega^{2s} (4+\hat m^2)^s {}_3F_2[\frac{1}{2},\frac{1-s}{2},-\frac{s}{2};1,1;\frac{16}{(\hat m^2+4)^2}]\,.
   \ee
   Using these expressions we can get the result for the sums involving $\log$ by using $f(s)=1/s^2$ in eq.(\ref{A1}) and for the linear $\hat \lambda$ by using $f(s)=1/(s+3/2)$ which essentially sets $s=-3/2$ in the arguments of the hypergeometric functions. For $d=2$, we can find explicit compact expressions. For the logarithmic case we get
   \be
   \sum_{i_1=0}^{N-1}\log\left(\frac{m^2+4\Omega^2\sin^2\left(\frac{\pi i_{1}}{N}\right)}{\tilde \omega_{ref}^2} \right)
     =2N \log \left(\frac{\sqrt{4+\hat m^2}+\hat m}{2\tilde\omega_{ref}\delta}\right)\,,
     \ee
     while for the interaction part we have
         
         \begin{eqnarray}
          \sum_{i_1=0}^{N-1}\frac{1}{\left(m^2+4 \Omega^2 \sin^2\left(\frac{\pi i_1}{N}\right)\right)^{3/2}}
         & =&\frac{N\Omega^{-3}}{(\hat m^2+2)^{3/2}}{}_2F_1[\frac{3}{4},\frac{5}{4},1,\frac{4}{(\hat m^2+2)^2}]\,,\\
         &=&\frac{2N \Omega^{-3}}{\hat m^2\pi (\hat m^2+4)^{1/2}}E[\frac{4}{\hat m^2+4}]\,,
          \end{eqnarray}
          where in the last line $E[k]$ is the elliptic function of the 2nd kind.
  For $d=3$, explicit expressions can be found as well although these are quite lengthy to quote here. Using these expressions we can expand around the lattice cutoff $\delta=1/\Omega$. For the logarithmic sum we find 
  \be
  N^2 \left(2\log(\frac{1}{\tilde \omega_{ref}\delta})+1.17+\hat m^2[0.30-0.16 \log(\hat m)]+O(\hat m^4)\right)\,,
  \ee
  while for the interacting sum  we get
  \be
  \frac{N^2\Omega^{-3}}{ (4+\hat m^2)^{3/2}} {}_3F_2[\frac{1}{2},\frac{3}{4},\frac{5}{4};1,1;\frac{16}{(4+\hat m^2)^2}]\,,
  \ee
  Expanding this around small $\delta$ we get as the leading term,
  \be
  \frac{N^2 \Omega^{-3}}{2\pi \hat m}+O(\hat m^0)\,.
  \ee
   The higher dimensional cases can presumably be done analogously yielding analytic results involving hypergeometric functions of the kind ${}_d F_{d-1}$. However, we will not attempt to give a general expression here. Using the $d=2,3$ examples we will resort to general expansions determining the coefficients numerically from the integral expressions.
    Given the analytical result for $d=2,3$ it is not hard to get the general result for arbitrary $d$ which takes the following form, 
\begin{align}
\begin{split} \label{gendsum}
&\frac{1}{4}\sum_{k=1}^{d-1}\sum_{i_k=0}^{N-1}\log\left(\frac{m^2+4\Omega^2\sum_{k=1}^{d-1}\sin^2\left(\frac{\pi i_{k}}{N}\right)}{\tilde \omega_{ref}^2} \right)\\&=\frac{V}{2\,\delta^{d-1}}\log\left(\frac{1}{ \tilde\omega_{ref}\, \delta}\right)+  \frac{V}{\delta^{d-1}}\left(a_{d-1}+\log(m\delta) \left[\sum_{k=0}b_k (m\delta)^k\right]+\sum_{k=1} c_k (m\delta)^k\right) .
\end{split}
\end{align}
   For the $\mathcal{O}(\lambda)$ term we proceed as follows. 

\begin{align}\label{intd4}
    \begin{split}
      \sum_{k=1}^{d-1}   \sum_{i_k=0}^{N-1}\frac{1}{\left(m^2+4 \Omega^2\sum_{k=1}^2\sin^2\left(\frac{\pi i}{N}\right)\right)^{3/2}}=&\frac{2 N^{d-1}}{\sqrt{\pi}}\int_{0}^{\infty}dt\sqrt{t} e^{-m^2 t} \left( e^{-2 t \Omega ^2} I_0\left(2 t
   \Omega ^2\right)\right){}^{d-1}
   \\=&\frac{2 N^{d-1} \delta^3}{\sqrt{\pi}}\int_{0}^{\infty}du \,  e^{-(m\delta) ^2 u} \sqrt{u}\left(e^{-2 u} I_0(2u)\right){}^{d-1}.
\end{split}
    \end{align}  \\
    In the second step we have made the change of variables to the dimensionless $u=t\, \Omega^2$; (where, $\Omega=1/\delta)$. We also notice that the the integral above is nothing but the Laplace transformation of the function $f(u)=\sqrt{u}\left( e^{-2 u} I_0(2u)\right){}^{d-1}$.  \\ \\  
  We will define $\mathcal{F}\{f(u)\}:=\int_{0}^{\infty}du  e^{-(m\delta) ^2 u} \sqrt{u}\left( e^{-2 u} I_0(2u)\right){}^{d-1}$. Now if we set $(m \delta)=0$ in  $\mathcal{F}\{f(u)\}$, then we observe that: $\mathcal{F}\{f(u)\}$ diverges for $d<4$, indicating a $1/(m \delta)^{n>0}$ behavior. A series expansion of the exact $d=2,3$ results suggest that the dependence on $m\delta$ is $(m \delta)^{d-4}$. Also from the plots shown in (...)  $\mathcal{F}\{f(u)\}$ vs $(m\delta)$,  for $d>4$ it can be easily seen  that it has a finite value at $(m\delta)=0$. These plots indicate that the fit for any d should behave as follows, 

\begin{align}
    \begin{split}\label{lapl}
        \mathcal{F}\{f(u)\}=& (f_1 (m \delta)^{d-4}+f_0); \hspace{1cm} \text{for } d\neq4\\
        =& (f_1 \log(m \delta)+f_0); \hspace{1cm} \text{for } d=4
         \end{split}
\end{align}

The methods developed in this appendix will prove useful for higher order perturbations, as well as other analogous sums.

\section{Structure of the Complexity beyond $\mathcal{O}(\lambda)$ }

Here, we will discuss higher order in $\lambda$  corrections to complexity focusing on the structure of the second block.
Let us first consider $\mathcal{O}(\lambda^2)$ correction to the target matrix. In addition to the correction to the first block, this introduces $O(\lambda^2)$ corrections to the second block and  a third block at $O(\lambda^2)$ comprising elements corresponding to coefficients of $$\tilde{x_a}^8, \tilde{x_b}^4 \tilde x_c^4, \tilde x_d^2 \tilde x_e^6, \tilde x_f^2 \tilde x_g^4 \tilde x_h^2, (\tilde x_i \tilde x_j \tilde x_k \tilde x_l)^2,$$ terms in the $\mathcal{O}(\lambda^2)$ block in the  target state and $O(\lambda^2)$ cross terms between the second and the third block due to coefficients of  $\tilde{x_a}^6, \tilde x_d^2 \tilde x_e^4,  (\tilde x_i \tilde x_j \tilde x_k )^2$ terms in the target state. In each block we only consider the leading order in $\lambda$. Henceforth, we focus on the $\mathcal{O}(\lambda^2)$ block  ($A_3$) and neglect these crossterms.

In $d=2$ dimensions, eigenvalues coming from this block ($A_3$)  takes the following  general form:
\begin{equation}
    \lambda_{k}^{(3)}= \left(\frac{\lambda}{N}\right)^2 \sum_{\substack{a,b=0\\\text{ Conditions on } a,b}}^{\text{Dim }\mathcal{A}_2-1}\frac{b_k }{f_{k}(\tilde\omega_a,\tilde\omega_{b})}.
\end{equation}\\ 
$k$ runs over the dimensions of $A_3$ and $f_{k}$'s are the quadratic polynomials of the  frequencies $\tilde \omega_{a}$ where $a,b=0,\cdots N-1.$  . Since we are only interested in the divergence structure of the leading contribution to complexity hence we ignore certain terms and consider the more simpler form for the eigenvalues given by,
\begin{equation}
    \lambda_{k}^{(3)}=\left(\frac{\lambda}{N}\right)^2\frac{b_k}{\tilde\omega_k^2}.
\end{equation}
In the general $O(\lambda^g)$ case we have $g+1$ blocks in the target matrix with the $(g+1)^{th}$ block containing coefficient of terms whose power of $\tilde x_i$ sums up to $4g$. At this order (order of $\lambda=g$) these terms are of purely $O(\lambda^g)$. The other diagonal blocks- say the $j^{th}$ block, j $\in$ $\{0,1,...,g\}$ have leading terms of $O(\lambda^{(j-1))}$ with subleading terms upto $O(\lambda^g)$. The cross terms between the $j^{th}$ block and the $(j+1)^{th}$ block will contain terms of at least $O(\lambda^j)$. In each diagonal block we consider only the leading term. With this  the eigenvalues of the general $j^{th}$ block is of the general form,
\begin{equation}
    \lambda_{k}^{j-1}= b_k\, \left(\frac{\lambda}{N}\right)^{j-1}\sum_{\substack{\alpha_1, \alpha_2,\alpha_3,..\alpha_{(k-1)},\alpha_{(k+1)},..\alpha_{(j-1)}=0\\\text{ Conditions on } \alpha_1,\alpha_2,..,\alpha_k,..,\alpha_{(j-1)}}}^{\text{Dim }\mathcal{A}_2-1}\frac{1}{ f_k(\tilde\omega_{\alpha_1},\tilde\omega_{\alpha_2},..,\tilde\omega_{\alpha_k},..\tilde\omega_{\alpha_{(j-1)}})}
\end{equation}
$f_k$'s in general a $j-1$ degree polynomial of the frequencies $\tilde \omega_{a},$ where $ a=0,\cdots N-1.$We consider $f(\tilde\omega_{\alpha_1},\tilde\omega_{\alpha_2},..,\tilde\omega_{\alpha_k},..\tilde\omega_{\alpha_{(j-1)}}) =\tilde\omega_{\alpha_1},\tilde\omega_{\alpha_2},..,\tilde\omega_{\alpha_k},..\tilde\omega_{\alpha_{(j-1)}}$. Since we are only interested in the contribution of the leading divergent term to complexity, therefore we ignore certain terms arising as a result of the conditional sum and consider the much more simpler set of eigenvalues given by,
\begin{equation}
    \lambda_{k}^{(j-1)}=  \left(\frac{\lambda}{N}\right)^{j-1}\frac{b_k}{ \tilde\omega_{k}^{j-1}}.
\end{equation}
We have counted that dimensions of each of these blocks are $N^2. $ Now as explained in the main text, each of these $N^2$ number of eigenvalues has a degeneracy of $N.$  This argument can be easily generalized for arbitrary dimensions $d.$  The contribution to the complexity coming from these blocks are given by (assuming the cost function $\mathcal{C}_{\kappa=1}$) 
\begin{align}
  \begin{split}  \label{ho1}
C_{\kappa=1}^{j}=&(j-1)\frac{V}{\delta^{d-1}}\frac{(\hat\lambda\, \delta^{4-d})^{(j-1)\mu} \delta^{-\nu}\, V^{\frac{\nu}{d-1}}}{2}\sum_{\{i_k\}=0}^{N-1}\log\left(\frac{b_{i_k}\, \hat \lambda\, \delta^{2-d}}{24\, L\, \lambda_0\, \tilde\omega_{i_{k}} \tilde\omega_{ref}}\right)\\
         =&(j-1) \frac{V}{\delta^{d-1}}\frac{(\hat\lambda\, \delta^{4-d})^{(j-1)\mu} \delta^{-\nu}\, V^{\frac{\nu}{d-1}}}{2}\Bigg\{\sum_{\{i_k\}=0}^{N-1}\log\left(\frac{b_{i_k}\,\hat\lambda\,  \delta^{2-d}}{24\, L\, \lambda_0}\right)
         \\&-\frac{1}{2}\sum_{\{i_k\}=0}^{N-1}\log\left(1+\frac{4\Omega^2}{m^2}\sum_{k=1}^{d-1}\sin^2\left(\frac{\pi\ i_k}{N}\right)\right)\Bigg\}.
\end{split}
\end{align}
We used the following form of the penalty factor,
\be 
\mathcal{A}^{j}= (\hat\lambda\, \delta^{4-d})^{(j-1)\,\mu} N^{\nu}.
\ee
Also, as explained in the main text we can chose for the reference,
\be
\lambda_0=a\hat\lambda \tilde\omega_{ref}^{d-3}, \text{or}\,  \lambda_0=a\hat\lambda m^{d-3}.
\ee
Then at each order $j,$ we can make either of the two choices as mentioned in (\ref{choice1}) and (\ref{choice2}). \\ The complexity and RG equations resulting from the general $j^{th}$ block seems to behave just like the $2^{nd}$ block, except for a factor of $(j-1)$.

     \end{appendix}

   \end{document}